\documentclass[a4paper,twocolumn,11pt, accepted=2023-01-03]{quantumarticle}
\pdfoutput=1

\usepackage[colorlinks=true,linkcolor=blue,urlcolor=blue,citecolor=blue,anchorcolor=green,pdfusetitle]{hyperref}
\usepackage{amsmath}
\usepackage{verbatim}
\usepackage{graphicx}
\usepackage{xcolor}
\usepackage{bbold}
\usepackage{amssymb}
\usepackage{amsthm}
\usepackage[braket,qm]{qcircuit}
\usepackage{physics}
\usepackage{comment}
\usepackage{tikz}
\usepackage{hyperref}
\usepackage{xspace}
\usepackage[utf8]{inputenc}

\usepackage[style=numeric-comp,
backend=bibtex8,
doi=true,
isbn=false,
url=false,
eprint=false,
maxbibnames=5,
sorting=none]{biblatex}
\renewbibmacro{in:}{}
\newcommand{\algo}{\mbox{AEQuO}\xspace}
\newcommand{\costF}{(\Delta\tilde{O})^2}
\newcommand{\sgn}{\text{sgn}}

\theoremstyle{plain}

\newcommand{\cov}{\text{cov}}

\newcommand{\HadProd}{\text{\hspace{0.1em}\large{$\circ$}\hspace{0.1em}}}
\newcommand{\HadDiv}{\raisebox{0.17em}{\text{\hspace{0.1em}\tiny{$\oslash$}\hspace{0.1em}}}}

\DeclareMathOperator{\Dir}{Dir}
\definecolor{green}{RGB}{0,128,0}

\begin{document}

\title{
Adaptive estimation of quantum observables
}
\author{Ariel Shlosberg}
    \email{ariel.shlosberg@colorado.edu}
    \thanks{this author contributed equally}
    \affiliation{JILA, University of Colorado and National Institute of Standards and Technology, Boulder, CO 80309, USA}
    \affiliation{Department of Physics, University of Colorado, Boulder, CO 80309, USA}
\author{Andrew J. Jena}
    \email{ajjena@uwaterloo.ca}
    \thanks{this author contributed equally}
    \affiliation{Institute for Quantum Computing, University of Waterloo, Waterloo, ON N2L 3G1, Canada}
    \affiliation{Department of Combinatorics \& Optimization, University of Waterloo, Waterloo, ON N2L 3G1, Canada}
\author{Priyanka Mukhopadhyay}
    \email{p3mukhop@uwaterloo.ca}
    \affiliation{Institute for Quantum Computing, University of Waterloo, Waterloo, ON N2L 3G1, Canada}
    \affiliation{Department of Combinatorics \& Optimization, University of Waterloo, Waterloo, ON N2L 3G1, Canada}
\author{Jan F. Haase}
    \email{jan.frhaase@gmail.com}
    \affiliation{Institute for Quantum Computing, University of Waterloo, Waterloo, ON N2L 3G1, Canada}
    \affiliation{Department of Physics \& Astronomy, University of Waterloo, Waterloo, ON N2L 3G1, Canada}
    \affiliation{Institute of Theoretical Physics and IQST, Universit{\"a}t Ulm, D-89069 Ulm, Germany}
\author{Felix Leditzky}
    \email{leditzky@illinois.edu}
    \affiliation{Institute for Quantum Computing, University of Waterloo, Waterloo, ON N2L 3G1, Canada}
    \affiliation{Department of Combinatorics \& Optimization, University of Waterloo, Waterloo, ON N2L 3G1, Canada}
    \affiliation{Department of Mathematics and IQUIST, University of Illinois Urbana-Champaign, Urbana, IL 61801, USA}
    \affiliation{Perimeter Institute for Theoretical Physics, Waterloo, ON N2L 2Y5, Canada}
\author{Luca Dellantonio}
    \email{luca.dellantonio@uwaterloo.ca}
    \affiliation{Institute for Quantum Computing, University of Waterloo, Waterloo, ON N2L 3G1, Canada}
    \affiliation{Department of Physics \& Astronomy, University of Waterloo, Waterloo, ON N2L 3G1, Canada}
    \affiliation{Department of Physics and Astronomy, University of Exeter, Stocker Road, Exeter EX4 4QL, United Kingdom}

\begin{abstract}
The accurate estimation of quantum observables is a critical task in science. With progress on the hardware, measuring a quantum system will become increasingly demanding, particularly for variational protocols that require extensive sampling. 
Here, we introduce a measurement scheme that adaptively modifies the estimator based on previously obtained data.
Our algorithm, which we call {\algo}, 
continuously monitors both the estimated average and the associated error of the considered observable, and determines the next measurement step based on this information. We allow both for overlap and non-bitwise commutation relations in the subsets of Pauli operators that are simultaneously probed, thereby maximizing the amount of gathered information.
{\algo} comes in two variants: a greedy bucket-filling algorithm with good performance for small problem instances, and a machine learning-based algorithm with more favorable scaling for larger instances.
The measurement configuration determined by these subroutines is further post-processed in order to lower the error on the estimator.
We test our protocol on chemistry Hamiltonians, for which
{\algo} provides error estimates that improve on all state-of-the-art methods based on various grouping techniques or randomized measurements, thus greatly lowering the toll of measurements in current and future quantum applications.
\end{abstract}
\maketitle
%

%
%

%
\section{Introduction}
\label{sec:introduction}
Quantum systems offer the prospect of accelerating computations \cite{Shor1994Algorithms, Nielsen2010, Acin2018The-quantum, Preskill2018Quantum} and simulations \cite{Georgescu2014Quantum} within a wide range of applications, such as high-energy physics \cite{Banuls2019Simulating,haase2021resource,paulson2020simulating} and chemistry \cite{Cao2019Quantum}.
Currently, so called noisy intermediate-scale quantum (NISQ) devices \cite{Preskill2018Quantum} provide a first flavor of the upcoming technology, but control errors \cite{Preskill2021Quantum} threaten the progress towards fault-tolerant, error-corrected hardware. Another challenge is the statistical nature of quantum mechanical measurements \cite{Breuer2002, 8585034,Itano1993}, requiring repeated measurements to estimate a quantum observable. 

The measurement problem is particularly relevant in current algorithms for NISQ devices that resort to an extensive sampling of the quantum system \cite{cerezo2021variational,ferguson2021}. The last few years have therefore seen an increased effort to find better measurement protocols that can lower the requirements on the quantum machine \cite{jena2019pauli,McClean2016The-theory,verteletskyi2020measurement,Arrasmith2020Operator,Crawford2021Efficient,Huang2021Efficient,Torlai2020Precise,hillmich2021decision,huang2020predicting,hadfield2020measurements,hadfield2021adaptive,Bujiao2021,kohda2021quantum,Gokhale2019,hamamura2020efficient,Yen2019Measuring,Izmaylov2019}. An optimal strategy to measure a quantum observable is, however, still unknown. 

In this work, we introduce a novel algorithm called {\algo} that maximises the information obtained from sampling the quantum system and learns from previous outcomes to improve the allocation of the remaining measurement budget. Compared to previous strategies, {\algo} is based on the ability of on-the-fly estimating not only the average of a given quantum observable, but also its error. This allows us to faithfully determine the precision of the estimated quantity, without the requirement of deriving error bounds \cite{Arrasmith2020Operator, Huang2021Efficient,hadfield2020measurements,hadfield2021adaptive}. Furthermore, instantaneous error knowledge also permits better allocation of the measurement budget and is at the core of the learning capabilities of our algorithm.


We focus on the problem of accurately estimating a given observable $O$. 
This is a typical challenge encountered for example in quantum-classical hybrid protocols such as the variational quantum eigensolver \cite{cerezo2021variational,ferguson2021}.
In particular, we want to estimate the expectation value of
\begin{equation} \label{eq:operator_def}
O = \sum_{j=1}^n c_j P_j,\quad\mathrm{i.e.}\quad \langle O \rangle   = \sum_{j=1}^n c_j \langle P_j \rangle,  
\end{equation}
using $M$ repeated preparations of the quantum state. 
Here, all $c_j$ are real constants, the Pauli string (PS) $P_j$ labels a tensor product of Pauli operators, and $\langle Q\rangle \equiv \langle Q\rangle_\rho = \tr(\rho Q)$ denotes the expectation value of an observable $Q$ with respect to the quantum state $\rho$. 

Devising a protocol that allocates the budget $M$ to minimise the estimation error is far from trivial. 
Several approaches have been proposed, including joint measurements of pairwise commuting operators \cite{McClean2016The-theory}, randomized and derandomized measurements \cite{Arrasmith2020Operator, Huang2021Efficient,hadfield2020measurements,hadfield2021adaptive}, grouping by weights \cite{Crawford2021Efficient}, neural networks \cite{Torlai2020Precise}, minimizing the number of measurement groups \cite{jena2019pauli,verteletskyi2020measurement,hadfield2020measurements,Bujiao2021,Gokhale2019,hamamura2020efficient,Yen2019Measuring,Izmaylov2019}, and a method based on maximum entropy and optimal transport \cite{rouze2021learning}. 
In this work, we introduce a measurement scheme for estimating observables that adaptively allocates the measurement budget based on previously collected data, allows for both non-bitwise commutation between PS and overlap in their grouping, and assesses both the average and variance of the observable $O$ in Eq.~\eqref{eq:operator_def}.

The article is structured as follows. In Sec.~\ref{sec:overview}, we provide an overview of our main results. In Sec.~\ref{sec:theory}, we introduce the necessary background for observable estimation.
We explain in detail the process of encoding an observable as in Eq.~\eqref{eq:operator_def} in a weighted graph, and we discuss the connection between the estimation task and clique covers of a weighted graph.
We explain {\algo}'s subroutines in Sec.~\ref{sec:algorithm} and provide numerical results using chemistry Hamiltonians as a benchmark in Sec.~\ref{sec:results}.
We conclude in Sec.~\ref{sec:conclusions} with some open problems and future directions of research.
The appendices contain additional information further explaining our methods and results.

\section{Overview of the main results}\label{sec:overview}

We introduce the Adaptive Estimator of Quantum Observables (\algo), a protocol designed to allocate the measurement budget in order to minimise the error affecting the estimate of an observable $O$ as in Eq.~\eqref{eq:operator_def}.
{\algo} iteratively changes the employed estimator based on the continuous inflow of new measurement data, it allows for both non-bitwise commutation relations between the PS and overlap in their grouping, and it yields estimates for both the average value and the error of $O$. 
It also employs Bayesian statistics, which permits the inclusion of prior knowledge and gives meaningful results even for small sample sizes. 
As a benchmark, we use {\algo} to estimate chemistry Hamiltonians, obtaining error estimates that improve on all state-of-the-art methods of estimating quantum observables.

A qualitative description of {\algo}'s subroutines is given below.
First, an operator $O$ as in Eq.~\eqref{eq:operator_def} is encoded in a (weighted) graph with vertex set $\lbrace c_j P_j\rbrace$ \cite{jena2019pauli}.
An edge is drawn between $P_i$ and $P_j$ whenever the two PS commute, and the corresponding edge weight, representing their covariance, is initialized (see Eq.~\eqref{eq:gen_error_cov} and Sec.~\ref{sec:algorithm}).
An estimate of $\langle O\rangle$ and its error can be directly calculated from this graph by assigning the vertices to groups of commuting operators that can be measured simultaneously. 
Crucially, we allow for overlap among these groups, that is, the operators $P_j$ in Eq.~\eqref{eq:operator_def} can belong to multiple groups, thus effectively increasing the amount of gathered information. 

The second step	of {\algo} consists in iteratively assigning a number of measurements to each of these groups until the budget $M$ is exhausted. Importantly, we develop a method (see Sec.~\ref{sec:theory}) to calculate the estimation error that {\algo} minimizes while allocating the $M$ shots.
We provide two different subroutines for this assignment.
One is a greedy ``bucket-filling'' (BF) algorithm that assigns the $M$ measurements one by one. 
This algorithm gives a good estimation error in small problem instances, but performs poorly for large problem sizes due to its greedy nature.
To remedy this, we developed another subroutine based on machine learning (ML) that requires fewer repetitions, as it allocates a fraction of the total budget $M$ to the most promising groups of commuting PS in each iteration.
As a result, the ML algorithm is faster for large values of $M$ and $n$, the latter being the number of PS in Eq.~\eqref{eq:operator_def}.

In a final step, we post-process the gathered data to obtain the estimator characterized by the smallest error based on the performed measurements. 
The post-processing utilizes the cumulative knowledge gathered during the measurement phase. It considers the contribution to the estimation error from each PS in different commuting groups, removing it if it is statistically likely to improve the error.

\section{Theory}
\label{sec:theory}

The PS $P_j$ in Eq.~\eqref{eq:operator_def} acts on $d$ qubits, $P_j=\bigotimes_{i=1}^d \sigma_i^{I_j(i)}$. The superscript $I_j(i)$ labels the element of the PS, i.e. $I_j(i) \in \lbrace0,x,y,z\rbrace$, with $\sigma^0$ denoting the identity and $\sigma^x,\sigma^y,\sigma^z$ the Pauli matrices. The subscript $i$ indicates the corresponding qubit on which a Pauli operator acts. A decomposition of an observable into PS as in Eq.~\eqref{eq:operator_def} has at most $n\leq 4^d$ terms; however, in most physical examples the number of PS with non-zero weight $c_j$ scales polynomially in $d$.

In an experiment, we collect $m_j$ measurement outcomes (``shots'') for each $P_j$ and take their average to obtain an estimate $\tilde{P}_j$ of $\langle P_j \rangle$, which subsequently allows us to estimate $\langle O \rangle$ through $\tilde{O}=\sum_{j}c_j\tilde{P}_j$. Considering that two commuting PS $P_j$ and $P_i$ can be measured simultaneously (i.e., in the same shot), an estimate for the error $\costF$ affecting $\tilde{O}$ is
\begin{subequations} \label{eq:gen_error}
  \begin{align}
     & \costF = \sum_{j,i=1}^n c_j c_i\, \mathcal{C} (\tilde{P}_j;\tilde{P}_i), \label{eq:gen_error_tot} \\
     &\mathcal{C} (\tilde{P}_j;\tilde{P}_i) = \tilde{\mathcal{Q}}_{ji} \frac{ m_{ji} + \delta(m_{ji})\delta(j-i)}{m_jm_i + \delta(m_j)\delta(m_i)} . \label{eq:gen_error_cov}
  \end{align}
\end{subequations}
Here, $\tilde{\mathcal{Q}}$ is the covariance matrix of all measured data, $m_{ji}$ denotes the number of shots where $P_j$ and $P_i$ have been sampled simultaneously, and $\delta$ is the delta function with $\delta(0) = 1$ and $\delta(x) = 0$ for $x\neq 0$. 
In the limit of many measurements, 
\begin{equation}
    \tilde{\mathcal{Q}}_{ji}\rightarrow \cov(P_j,P_i) \equiv \expval{P_jP_i}-\expval{P_j}\expval{P_i}. \nonumber
\end{equation}
Note that $m_{jj} = m_j$, and that $\sum_j m_j = M$ if and only if $m_{ji}=0$ for all $j\not=i$. In this special case, Eqs.~\eqref{eq:gen_error} yield the well known result 
\begin{equation}
    \costF = \sum_{j}\frac{c_j^2(\Delta\tilde{P}_j)^2}{m_j}, \nonumber
\end{equation}
where $(\Delta\tilde{P}_j)^2 \equiv \tilde{\mathcal{Q}}_{jj}$.

\begin{figure}
	\centering
	\includegraphics[width=\columnwidth]{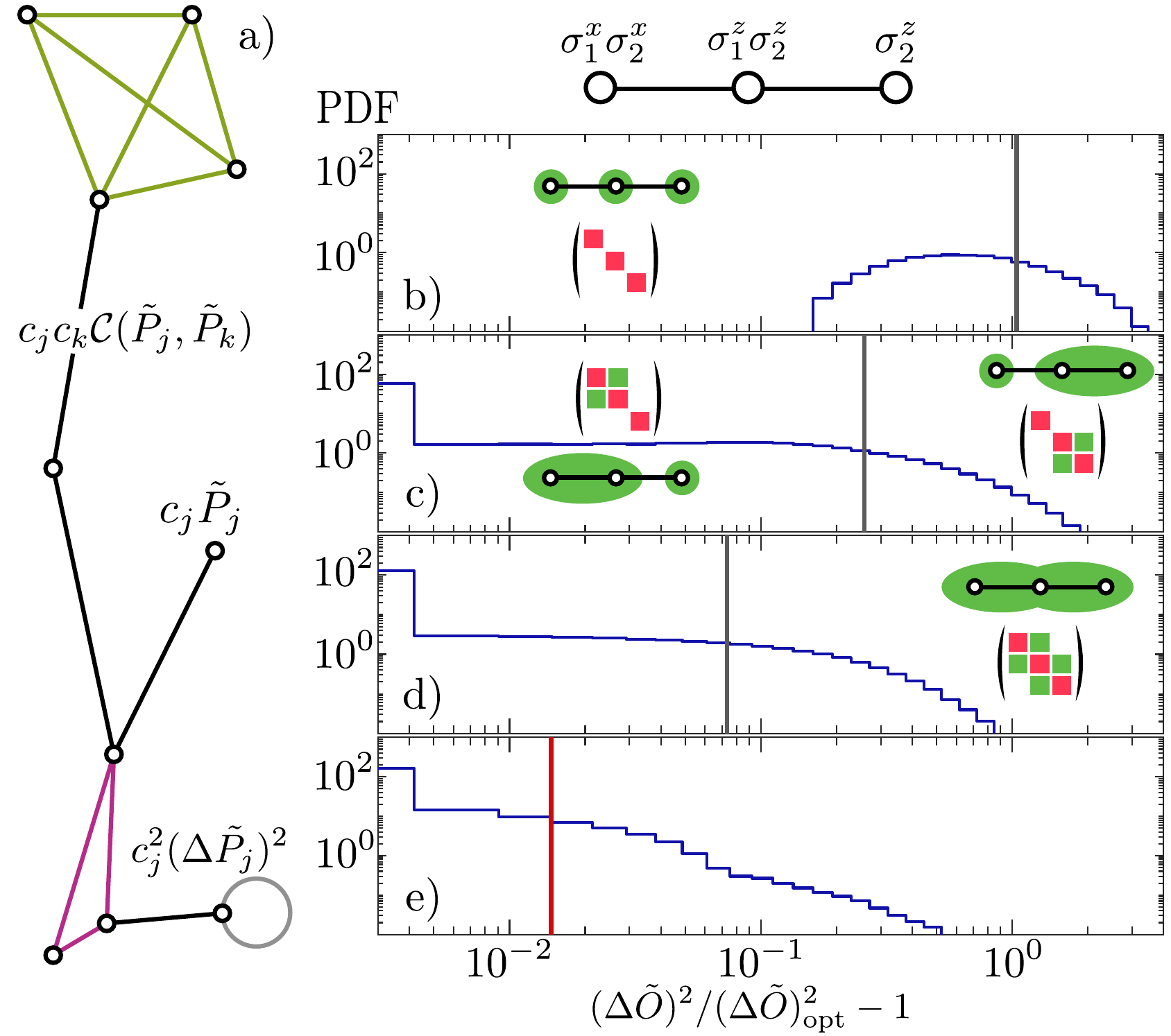}
	\caption{
	(a): Example of a graph. Vertices and edges are weighted with $c_{j}\tilde{P}_j$ and $c_{j}c_{i}\mathcal{C}(\tilde{P}_j;\tilde{P}_i)$, respectively. As shown by the grey circle, we include self-edges (omitted in the rest of the graph) with variances $c_j^2(\Delta\tilde{P}_j)^2$. Two maximal cliques are highlighted in green and violet. (b-e): Case study of the operator $O = \sigma_{1}^{x}\sigma_{2}^{x} + \sigma_{1}^{z}\sigma_{2}^{z} + \sigma_{2}^{z} $ using its graph representation (top). The four non redundant covers 
	are highlighted within plots (b-d), along with the $\tilde{\mathcal{Q}}_{ji}$ elements contributing to the associated error $\costF$ (matrices with red and green squares). Results from measurements of these covers are plotted in the graphs. We considered $2\cdot 10^5$ experiments where $M = 10^4$ experimental shots are used to estimate $\tilde{O}$ and $\costF$ for pure states uniformly sampled on the Bloch sphere. For each cover, measurement allocation is optimal and is determined with the full knowledge of $\tilde{\mathcal{Q}}$. As discussed in Sec.~\ref{sec:theory} and App.~\ref{app:overlap}, for each state one cover gives the smallest error $\costF_{\rm opt}$, to which the results obtained for the four plots are compared. Vertical lines represent averages. In the bottom panel (e), we collect the results obtained by {\algo} \textit{without} prior knowledge of $\tilde{\mathcal{Q}}$, and using the ML subroutine with $L=1$. 
	}
	\label{fig:Fig1}
\end{figure}

To obtain $\tilde{O}$, there is no unique way to choose groups of PS that are pairwise commuting, in particular if one allows these groups to have overlap. Every choice of groups corresponds to a specific estimator in the form of Eqs.~\eqref{eq:gen_error} characterized by the values of $m_{ji}$ and $m_j$. 
In the following, we explain the aforementioned graph representation used in our protocol, and the idea behind our algorithm for grouping PS.

As shown in Fig.~\ref{fig:Fig1}(a), we represent the operator $O$ in Eq.~\eqref{eq:operator_def} with a weighted graph whose vertices correspond to $c_j\tilde{P}_j$. If $P_j$ and $P_i$ commute, their vertices are connected by an edge with weight $c_jc_i\,\mathcal{C}(\tilde{P}_j,\tilde{P}_i)$. Therefore, $\tilde{O}$ is obtained by summing over all vertices, while $\costF$ is the sum of all edges (including self-edges). Different $P_j$ can only be measured simultaneously if they belong to the same clique of the graph, i.e., a fully connected subset of vertices. 
A clique is called maximal if no further vertex can be added to it [see Fig.~\ref{fig:Fig1}(a)]. Evidently, an estimator corresponds to a clique cover, i.e., a set of cliques such that each vertex of the graph is included in at least one clique.

For illustrative purpose, let us consider the example $O = \sigma_{1}^{x}\sigma_{2}^{x} + \sigma_{1}^{z}\sigma_{2}^{z} + \sigma_{2}^{z}$ in Fig.~\ref{fig:Fig1}(b-e). There are five cliques that can be arranged in four different, non-redundant covers\footnote{In general, one can find more covers by nesting smaller cliques into bigger ones. However, these cases are not relevant to us since only one of these nested cliques will yield minimal error. Therefore, without loss of generality, we can limit ourselves to the study of the four covers in Fig.~\ref{fig:Fig1}.}. The particular choice of clique cover determines the terms $\mathcal{C}(\tilde{P}_j;\tilde{P}_i)$ contributing to the error $\costF$, as can be seen in the pictorial representations of the matrices $\tilde{\mathcal{Q}}$ in Fig.~\ref{fig:Fig1}(b-d). For a given input state, the estimator $\tilde{O}$ corresponding to the cover yielding the minimal $\costF$ for fixed $M$ gives the most accurate estimation (since all estimators are unbiased). 

We test the performances of the four estimators (viz.~covers) in Fig.~\ref{fig:Fig1}(b-d) with randomly chosen two-qubit pure states, and calculate the relative deviation of the error estimate w.r.t.~the best estimator. 
Here, we use the asymptotic values for Eq.~\eqref{eq:gen_error_cov} that are obtained for $M\rightarrow \infty$.
The results demonstrate that, as expected, measuring each PS separately [Fig.~\ref{fig:Fig1}(b)] is never a good strategy, while the intuitively best estimator [Fig.~\ref{fig:Fig1}(d)], corresponding to the cover made of maximal cliques only, is optimal in only ${\sim}54\%$ of the cases. More details about how the best estimator is determined and the measurement budget is allocated can be found in App.~\ref{app:overlap}.

In practice, the state to be measured is unknown and it is not possible to test all covers beforehand to identify the best one. This motivates an adaptive approach in which the clique cover is changed during the measurement process depending on former outcomes. In our protocol, we update the vertex and edge weights of the graph after some (or each) of the $M$ shots. We then decide which clique is measured next, based on its contribution to Eq.~\eqref{eq:gen_error_tot}.

Conveniently, the choice of the particular estimator can be relegated to a second post-processing step; in the data acquisition phase we may restrict to considering maximal cliques only. Changing the estimator (or equivalently, the clique cover) can then be done subsequently by adjusting the measurement numbers $m_{ji}$ and $m_j$, and updating $\tilde{Q}$. For example, if one removes $P_j$ from a clique, then for all $P_i$ in that clique one has to reduce $m_{ji}$ and $m_j$ by the number of shots allocated to that clique. Furthermore, all outcomes gathered for $P_j$ in the considered clique have to be discarded in order to avoid introducing biases in the estimator. In the example in Fig.~\ref{fig:Fig1}(d), changing the estimator corresponds to removing the PS $\sigma_{1}^{z}\sigma_{2}^{z}$ from one of the two cliques.

The advantages of employing maximal cliques and post-processing are twofold. First, restricting to maximal cliques reduces the number of choices in each shot.
Second, potentially available data is never neglected, resulting in a better knowledge of the vertex and edge weights of the graph. This knowledge can then be used for more informed choices, both of the maximal cliques to be measured at each shot, and in the post-processing stage itself.

In the histograms of Fig.~\ref{fig:Fig1}(b-e), we plot the expected deviations from the minimal error, which is derived by always identifying the best cover out of the four. 
The black and red lines in the figures indicate the mean values of the associated histograms.
A comparison shows that {\algo} [in Fig.~\ref{fig:Fig1}(e)], based on maximal cliques and post-processing, consistently provides near-minimal error and greatly outperforms the static approaches [in Fig.~\ref{fig:Fig1}(b-d)] represented by the four covers in the figure. For more details, see App.~\ref{app:overlap}.

As a conclusive remark, we present two corollaries following from our graph representation of the observable $O$ and Eqs.~\eqref{eq:gen_error}. Namely, we derive simple upper bounds on both the error $\costF$ and its scaling with respect to the number $n$ of PS within $O$. To bound the error, we observe that the maximum error contribution of two PS $P_j$ and $P_i$ measured together is $c_j^2 + c_i^2 + 2 \lvert c_j c_i \rvert$,
which can be obtained by setting all $\tilde{\mathcal{Q}}_{ji}$ equal to $\sgn(c_j)\sgn(c_i)$  in Eq.~\eqref{eq:gen_error_cov}.
This yields the tight upper bound 
%
\begin{equation}
    \costF \leq \sum_{j,i=1}^n\left\lvert c_j c_i \right\rvert \frac{m_{ji}}{m_j m_i} .\nonumber
\end{equation}
Here, we have omitted the delta functions for clarity. Given any measurement strategy, this equation determines, \textit{beforehand}, the maximum error that can possibly affect the resulting estimator. The scaling of $\costF$ in terms of $n$ can be understood using graph theory. Considering that it is always possible to find a graph's cover of $n$ cliques, one concludes that in the worst-case scenario $\costF$ grows linearly in $n$.

\section{Algorithm}
\label{sec:algorithm}
\begin{figure}
    \centering
    \includegraphics[width=0.95\columnwidth]{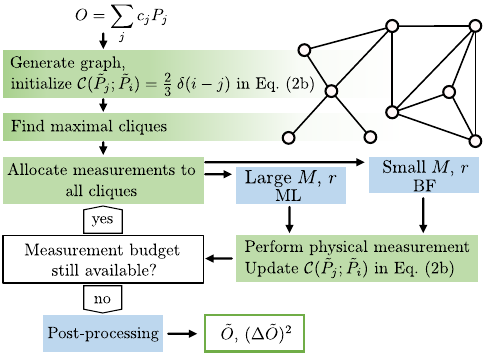}
    \caption{
    Schematic diagram of our protocol. From a given observable $O$, its graph representation is derived and the maximal cliques are found. Depending on the number of cliques $r$ and the measurement budget $M$, the BF or the ML algorithm is chosen. Afterwards, the system is probed and vertices and edges of the graph are updated according to the outcomes. If $M$ shots have been taken, the post-processing resorts to the desired estimates for the average and error. Otherwise, another round of allocation and measurements is performed. Green and light blue boxes represents essential and optional subroutines, respectively. In fact, the user is free to choose between the BF and the ML algorithm, while post-processing can be turned off if desired.}
    \label{fig:Fig2}
\end{figure}

In the following, we present our algorithm {\algo}, which is outlined in the diagram in Fig.~\ref{fig:Fig2}. First, the graph is constructed in time $\mathcal{O}(n^2)$. At this stage, it is possible to choose whether or not to connect vertices whose associated PS are non-bitwise commuting. As discussed below in Sec.~\ref{sec:results}, restricting to bitwise commuting PS results in errors $\costF$ that are higher. However, simultaneously measuring non-bitwise commuting PS requires entangling gates, which in the context of NISQ devices are expensive. {\algo} includes a sub-routine for finding a suitable circuit that diagonalizes a group of commuting PS \cite{github,jena2019pauli,improved_sim}. This subroutine is classically efficient for large values of the numbers $n$ and $d$ of PS and qubits, respectively, and the resulting circuit is ensured to contain at most $d(d-1)/2$ (zero) entangling gates when allowing (forbidding) non-bitwise commutation.

After the graph is built, {\algo} finds a desired number $r$ of maximal cliques. We prioritize cliques that are statistically likely to have bigger contributions to the error $\costF$ in Eq.~\eqref{eq:gen_error_tot} (see App.~\ref{app:our_algorithm} and \cite{github}) and only consider maximal ones for the reasons explained above. However, our protocol may also operate on all cliques. The runtime required to find the clique cover scales as $\mathcal{O}(r)$.
Efficient algorithms for finding maximal cliques are known, e.g., the Bron-Kerbosch algorithm \cite{bron1973algorithm}.

We now assign weights to the vertices and edges of the graph constructed from the observable $O$. We resort to Bayesian estimation (see App.~\ref{app:bayes}) for both the initialization and all subsequent updates of the weights of the graph, which gives meaningful results with scarce statistics or even in the absence of any data. This is crucial due to the adaptive nature of our protocol, and allows us to initialize the graph without any pre-sampling. Indeed, when no shots are taken, Bayesian estimation prescribes $\tilde{P}_j = 0$ and $\tilde{\mathcal{Q}}_{ji} = 2\delta(i-j)/3$ for all $j,i=1,\dots,n$.

Once the $r$ cliques have been found and the graph has been initialized, the system can be measured. 
In order to choose the clique to be probed at each shot, we propose two possible subroutines whose objective is to minimize the cost function $\costF$ in Eqs.~\eqref{eq:gen_error} (for a numerically efficient method to compute $\costF$ see App.~\ref{app:variance_calc}).
The first choice is a greedy ``bucket-filling'' (BF) algorithm \cite{cormen2009introduction} whose premise is to allocate measurements one-by-one by evaluating the predicted cost function before each shot. After a certain amount of measurements (the $M$ shots are divided into $L$ chunks of increasing sizes, as explained below), the graph is updated and the BF algorithm is run again. This approach works well for instances where $r$ and $M$ are small, since the number of cost function evaluations increases linearly in these variables.

As an alternative to the BF method we develop an approach inspired by a recent structural optimization algorithm based on machine learning (ML) \cite{hoyer2019neural}. We reparamterize the vector of clique measurements, which are the set of optimization variables in the BF approach, in terms of the weights and biases of a densely-connected neural network. The output of the network yields the (locally) optimal measurement allocations after learning occurs. This approach alters the optimization landscape by increasing the number of parameters.

In contrast to BF, the ML algorithm allocates a fraction of the total budget to the most promising cliques. Subsequently, the graph is updated and the ML algorithm is run again. We perform $L$ iterations of the process, at each step updating the covariance matrix $\tilde{\mathcal{Q}}$ with the experimental outcomes and progressively increasing the available shots by a factor $l > 0$ until all $M$ measurements are exhausted. Thus, for large problem sizes the ML subroutine scales more favorably compared to the BF algorithm, since it requires fewer repetitions ($L$ for ML versus $M$ for BF), and is more efficient for large values of $r$.

The ML approach is based on minimizing the cost function, Eq.~\eqref{eq:gen_error_tot}, over the trainable parameters of the neural network. We implement a variety of different optimizers to determine the weights and activation biases \cite{github}, including stochastic gradient descent \cite{robbins1951stochastic}, Adam \cite{Kingma:2014aa}, and a Limited-Memory BFGS method \cite{wright1999numerical,robbins1951stochastic,gill1972quasi}. These different optimizers yield similar performances; for instance, for Figs.~\ref{fig:Fig3} and \ref{fig:Fig4} the stochastic gradient descent and the Limited-Memory BFGS methods, respectively, were employed. 

The architecture of the neural network we settled on is composed of three densely-connected layers of width $r$, and with ReLU activation functions on hidden layers and a softmax activation function on the output \cite{Nwankpa2018Activation}. Such a structure yields a probability distribution at the output corresponding to the ratio of times that each clique should be measured. The output of the neural network is then converted to an integer measurement allocation vector by first scaling by the number of measurements to be performed, then flooring the entries, and finally by allocating excess measurements to the cliques with the largest percentage change in budget due to the flooring operation \cite{github}. Before the learning phase, the network is initialized such that each clique is measured the same number of times and so that measurements between various cliques are initially uncorrelated.


After completing all measurements, post-processing can be applied. In this step of the protocol, we determine the estimator with the lowest variance that could be realized with the available data. For each pair $P_j$ and $P_i$ such that $m_{ji} \neq 0$ and $c_j c_i \mathcal{C}(\tilde{P}_j;\tilde{P}_i) > 0$, we find all cliques where these strings have been simultaneously probed and consider all possibilities of removing either of those or keeping the estimator as it is. Eventually, the configuration minimizing the updated error function $\costF$ is kept.
The runtime of this procedure scales exponentially in the size of the subset of cliques where $P_j$ and $P_i$ have been measured simultaneously. In practical examples, this number is small and post-processing can be used for large values of $n$. For instance, post-processing has been used for all numerical results presented in Fig.~\ref{fig:Fig3}, where we considered observables with up to $n=1086$ terms corresponding to the Hamiltonian of $\rm{H}_2 \rm{O}$.

\section{Results}
\label{sec:results}

\begin{figure}
    \centering
    \includegraphics[width=\columnwidth]{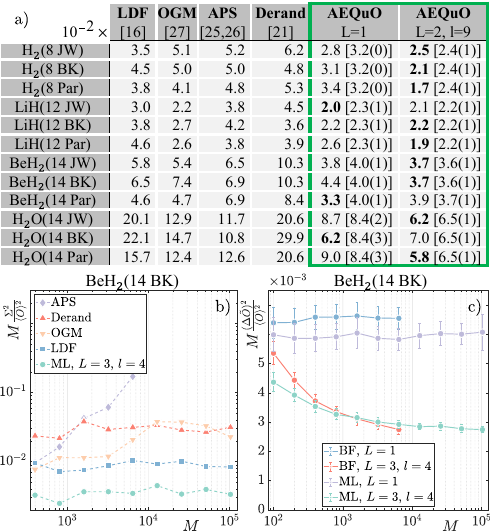}
    \caption{
    (a): Errors obtained with LDF \cite{jena2019pauli}, OGM \cite{Bujiao2021}, APS \cite{hadfield2020measurements,hadfield2021adaptive}, Derand \cite{Huang2021Efficient}, and {\algo} with the ML subroutine and $L=1$ or $L=2$ and $l=9$. We consider chemistry Hamiltonians with different numbers of qubits (in parentheses), and three encodings: Jordan-Wigner (JW), Bravyi-Kitaev (BK), and parity (Par) \cite{Cao2019Quantum}. Values (in bold the lowest) are standard deviations $\Sigma \equiv [\sum_{j=1}^R(\tilde{O}_j - \langle O \rangle)^2/R]^{1/2}$, where $\tilde{O}_j$ is the $j$-th estimated average for $O$. In square brackets, we report averages of $[\costF]^{1/2}$ 
    with their statistical error. The input is the ground state, $R=40$, $M=10^3$, and values are rescaled by $10^{-2}$. 
    (b-c): Relative errors 
    as a function of $M$ obtained considering the $\rm{BeH}_2$ Hamiltonian with BK encoding. In (b) we compare values of $M \Sigma^2/\langle O \rangle^2$ for the approaches in the legend.
    In (c), we report $M \costF / \langle O \rangle^2$ calculated with the BF and ML subroutines of {\algo} with different $L$ and $l$.
    Error bars are statistical errors. In (b) and (c) we set $R=10^2$ and $R=40$, respectively. Details in App.~\ref{app:our_algorithm}.}
    \label{fig:Fig3}
\end{figure}

Our numerical results are reported in Fig.~\ref{fig:Fig3}. In panel (a), we list the standard deviations $\Sigma$ (see caption and App.~\ref{app:our_algorithm}) affecting the estimated ground state energies for chemistry Hamiltonians \cite{Cao2019Quantum} of various molecules, choosing $M=1000$ in each case. Values in square brackets are averages of $\sqrt{\costF}$, calculated by {\algo} with the graph representation described in Sec.~\ref{sec:theory}. We compare {\algo} with an in-house developed version (see below and App.~\ref{app:our_algorithm}) of the largest degree first (LDF) \cite{jena2019pauli}, the overlapped grouping measurement (OGM) \cite{Bujiao2021}, the adaptive Pauli shadow (APS) \cite{hadfield2020measurements,hadfield2021adaptive}, and the derandomized shadow (Derand) \cite{Huang2021Efficient} methods.

In the first two, the goal is to minimize the number of cliques in the resulting cover, while the latter two reconstruct the desired estimator from the outcomes of partially random as well as deterministically allocated measurements. We choose these as representatives of the variety of algorithms based on different grouping strategies \cite{jena2019pauli,McClean2016The-theory,verteletskyi2020measurement,Arrasmith2020Operator,Crawford2021Efficient,Bujiao2021,hamamura2020efficient,Yen2019Measuring,Izmaylov2019} and the classical shadow technique \cite{Huang2021Efficient,Torlai2020Precise,hillmich2021decision,huang2020predicting,hadfield2020measurements,hadfield2021adaptive}.
{\algo} is run both for $L=1$ and $L=2$ (in this case $l=9$) using the ML subroutine, and outperforms other approaches in determining more precise estimates. 

As discussed in more detail below, this improvement has two reasons. First, we allow for both non-bitwise commutation relations between PS and overlapping cliques. On the one hand, this allows us to gather more information from the same number of shots. On the other hand, it increases the number of non-zero elements $\tilde{\mathcal{Q}}_{ji}$ in Eqs.~\eqref{eq:gen_error}. While these covariances can either lower or increase the error $\costF$, the adaptive nature of {\algo} (see next paragraphs) and the post-processing ensure that negative ones are preferred. We remark that our version of the LDF protocol also allows for non-bitwise commutation relations; as a result, it typically outperforms OGM, APS and Derand. 
As shown in App.~\ref{app:bitwise_comp}, when restricted to bitwise commutation relations, LDF yields errors that are generally higher than all other protocols.

Second, {\algo} allocates the shots by directly minimizing the estimated error $\costF$ in Eqs.~\eqref{eq:gen_error} based on on the available experimental information. Provided the ML and/or the BF subroutines find the global minimum of $\costF$, this results in {\algo} finding the estimator $\tilde{O}$ that, constrained by the used cliques and the transient knowledge about $\tilde{\mathcal{Q}}$, is characterized by the smallest possible error. This is supported by Fig.~\ref{fig:Fig3}(b-c), where we plot the relative errors $M \Sigma^2 / \langle O \rangle^2$ and $M \costF / \langle O \rangle^2$, respectively, of the ${\rm BeH}_2$ Hamiltonian in the Bravyi-Kitaev (BK) encoding. Despite statistical fluctuations, panel (b) shows that {\algo} outperforms all other approaches. Similar to other methods, we also see that {\algo} provides errors that scale as $1/M$ in the chosen range of $M$.

\begin{figure}
    \centering
    \includegraphics[width=\columnwidth]{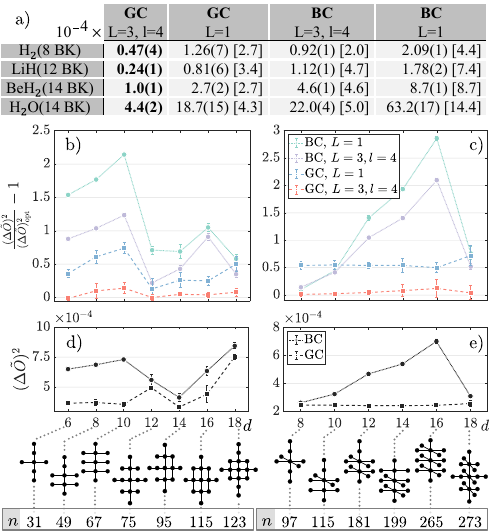}
    \caption{
    (a): Averaged errors $\costF$ (standard deviations in brackets, bold indicates lowest) obtained by {\algo} for chemistry Hamiltonians and different values of $L$ and $l$. BC (GC) refers to bitwise (general, i.e., non-bitwise) commutation. Relative overheads compared to the GC with $L=3$, $l=4$ are in square brackets.
    (b-c): Averaged relative errors $\costF / \costF_{\rm opt} - 1$ as a function of the number of qubits $d$ for different $2$D (b) and $3$D (c) lattices (see main text and App.~\ref{app:lattice_model}). Data are obtained with different values of $L$ and $l$, and resorting to BC or GC. 
    $\costF_{\rm opt}$ corresponds to the black squares connected by dashed lines in panels (d-e), where we report averaged errors $\costF$ obtained with prior knowledge of the state. For (b-e), the considered lattices with the associated $n$ are at the bottom, error bars are standard deviations and we used the ground states of the models. For all panels, $M=10^4$ and values are obtained by averaging $25$ independent runs of {\algo}.}
    \label{fig:Fig4}
\end{figure}

The adaptive features of {\algo} are evident in Fig.~\ref{fig:Fig3}(c), where we compare the BF and ML subroutines for different values of $L$ and $l$. For $L=1$ there is no memory of prior outcomes; in this case, the points obtained with either subroutine lie on a horizontal line. However, for $L=3$ and $l=4$, {\algo} learns and uses the gathered information to find better strategies for allocating the remaining shots. 
For large $M$, asymptotes are recovered, which are considerably lower than the ones for $L=1$. Importantly, the advantage from the adaptive nature of {\algo} comes for 
free, in the sense that it follows exclusively from a better allocation of the measurement budget and does not require extra resources (such as entangling gates for the simultaneous diagonalization of the PS). We investigate the improvement resulting from the adaptivity of {\algo} in more detail in Fig.~\ref{fig:Fig4} and in App.~\ref{app:bitwise_comp}, where we restrict all protocols considered in this section to bitwise commutation relations.

While the BF and ML subroutines yield comparable results, Fig.~\ref{fig:Fig3}(c) indicates that they outperform each other in different parameter regimes, depending on the problem at hand. This is a consequence of several factors. By construction, the BF picks the clique to be measured based on the gradient of $\costF$. On the other hand, the ML subroutine (in this context) follows a stochastic gradient descent algorithm \cite{robbins1951stochastic} that allows for exploring the error landscape. However, its performance depends on hyperparameters \cite{github,hoyer2019neural,Kingma:2014aa} that, in Fig.~\ref{fig:Fig3}, have not been optimized to keep low computational requirements. 

As explained above, our protocol has three distinctive features.
Besides the ability to monitor and determine the error $\costF$ [see Eqs.~\eqref{eq:gen_error}] that allows for better allocation of the measurements and assessment of the precision of the estimator, {\algo} can also exploit both non-bitwise commutation between PS\footnote{This ability is native to our LDF protocol as well.} and cliques' overlaps, and it is adaptive.
In the remainder of this section, we investigate the advantages provided by these last two features both with chemistry Hamiltonians and a family of $2$D and $3$D lattice models of interacting spin $1/2$ particles (for details, see App.~\ref{app:lattice_model}). 

In Fig.~\ref{fig:Fig4}(a), we list averaged errors $\costF$ for several chemistry Hamiltonians\footnote{We show results relative to the BK encoding; the JW and Parity ones are qualitatively identical.}, with $M=10^4$ and the ground state as input~\cite{Cao2019Quantum}.  {\algo} is run with different settings, namely (not) allowing non-bitwise commutation relations between the PS, and with adaptive allocation turned either on ($L=3$, $l=4$) or off ($L=1$). As expected, the smallest errors are obtained when {\algo} can learn from previous outcomes ($L=3$, $l=4$) and non-bitwise commuting PS can be simultaneously measured [indicated by GC in Fig.~\ref{fig:Fig4}(a)]. In the case of chemistry Hamiltonians with their ground state as input, the advantage is remarkable. If we compare {\algo} when restricted to bitwise commutation (BC) and $L=1$ with the best results (bold numbers), we see that measurement budgets are required to be up to $14.4$ times bigger [see numbers in square brackets in Fig.~\ref{fig:Fig4}(a)] in order to obtain the same estimator's precision. 

For comparison, up to statistical noise, the Derand \cite{Huang2021Efficient} and OGM \cite{Bujiao2021} protocols yield results that are similar to {\algo}'s when we restrict it to bitwise commutation and turn the adaptive allocation off (BC, $L=1$). The direct comparison between the BC versions of {\algo} and all other protocols considered in this section is presented in App.~\ref{app:bitwise_comp}.

A similar analysis is made in Fig.~\ref{fig:Fig4}(b-e), where we considered a generalized Hubbard model \cite{essler2005one} in which qubits are located on edges of the $2$D [panels (b) and (d)] and $3$D [panels (c) and (e)] lattices drawn at the bottom of the figure. As explained in App.~\ref{app:lattice_model}, they are allowed to hop with (many-body) flip-flop interactions, and we include energy shifts that depend on the states of qubits on neighbouring edges. We use the ground state as input, set $M=10^4$, and run {\algo} with the same settings as in Fig.~\ref{fig:Fig4}(a): allowing for (squares) or forbidding (circles) non-bitwise commutation, and turning the adaptive allocation on (red and violet points) or off (blue and green points).

In panels (b-c), we show the relative increase of the error $\costF$ if compared to $\costF_{\rm opt}$, that is the value obtained by {\algo} when allowing for non-bitwise commutation and providing prior knowledge of the input state. In practice, prior knowledge is given by initializing the covariance matrix $\tilde{Q}_{ji}$ in Eq.~\eqref{eq:gen_error_cov} with $\cov(P_j,P_i)$ for all $i,j=1,\dots,n$. Therefore, the data points in Fig.~\ref{fig:Fig4}(b-c) represent the relative increase of the measurement budget $M$ that is required for reaching the error $\costF_{\rm opt}$. Values for $\costF_{\rm opt}$ correspond to the black squares in panels (d-e), depicting the errors $\costF$ determined by {\algo} with prior knowledge of the state.

It is evident from Fig.~\ref{fig:Fig4} that the advantages from the adaptive allocation and non-bitwise commutation are consistent, yet vary considerably between data points. All red squares in panels (b) and (d) are approximately zero, indicating that {\algo} learns in the process and for sufficiently large $M$ it allocates the measurements \textit{as if} it knew the input state \textit{beforehand}. All other coloured data points lie above, suggesting that it is detrimental not to exploit non-bitwise commutation and adaptive allocation.

How advantageous these features are \textit{highly} depends on the considered observable and state. Empirically, non-bitwise commutation is more beneficial if the Hamiltonian is dominated by long-range many-qubits interactions, such that two commuting PS are more likely to commute non-bitwise. Chemistry Hamiltonians fall within this category. On the other hand, in most of the considered examples adaptive allocation resorted to precision improvements varying between $25\%$ and $400\%$, with the tendency of increasing with the numbers of qubits $d$ and PS $n$.


%
%
\section{Conclusions and outlook}
\label{sec:conclusions}

We introduced {\algo}, an adaptive algorithm to estimate quantum observables expressed as a sum of Pauli operators.
%
%
%
By allowing for both overlap between groups of PS that are simultaneously measured and general commutation relations, we gather more information per shot and introduce possibly negative covariances into Eqs.~\eqref{eq:gen_error}. {\algo} is capable of allocating the remaining measurement budget depending on previous outcomes, it post-processes the estimator to lower its error, and it gives estimates both for the average $\tilde{O}$ and the variance $\costF$ of the considered observable. Our protocol is based on two routines which provide either near-optimal (BF) or computationally efficient (ML) allocation of the measurement budget. For the observables considered here, both subroutines result in similar errors $\costF$.

We tested our algorithm on several Hamiltonians with different settings that include allowing or forbidding non-bitwise commutation between PS, and turning the adaptive allocation on ($L>1$) or off ($L=1$). 
Our protocol yields numerical results that outperform state-of-the-art estimators based on various grouping techniques \cite{jena2019pauli,verteletskyi2020measurement,hadfield2020measurements,Bujiao2021} and the classical shadow method \cite{hadfield2020measurements,Huang2021Efficient,hadfield2021adaptive}. We also studied the advantages our algorithm gains from adaptive allocation as well as non-bitwise commutation. We found that, while being highly problem dependent, these advantages are consistent.

There are different possibilities for improving our algorithm.
Generating the list of all maximal cliques of the graph is computationally demanding for large problem instances. In this case, one could find better strategies to select a subset of (maximal) cliques, and this subset can in principle be updated while measurements are taken. 
Designing a better method of choosing suitable cliques could increase the performance of our algorithm without sacrificing the quality of the results. 
The ML-based subroutine could be further improved by employing  graph neural networks to leverage the graph structure of the problem \cite{graphneuralnetworks}.
Another possibility is the extension of our adaptive algorithm to non-qubit-based hardware and, in view of the Bayesian framework underlying our estimation, the direct inclusion of the experimental characteristics of the considered platform via hierarchical modelling \cite{Haase2018}.

\section*{Acknowledgements} 
We thank Johannes Bausch, Michele Mosca, Graeme Smith,  and Peter Zoller for fruitful and enlightening discussions, and Hsin-Yuan Huang and Charles Hadfield for help with the Derand \cite{huang2020predicting} and the APS \cite{hadfield2020measurements,hadfield2021adaptive} codes, respectively. We are also grateful to Jinglei Zhang for valuable feedback on a prior version of this manuscript.
This work has been supported by Transformative Quantum Technologies Program (CFREF), NSERC, PWGSC and the New Frontiers in Research Fund. Research at IQC is supported in part by the Government of Canada through ISED.
JFH acknowledges the Alexander von Humboldt Foundation in the form of a Feodor Lynen Fellowship, the ERC Synergy Grant HyperQ (grant number 856432), the EU Flagship project AsteriQs (GA-820394-ASTERIQS), and the BMBF project Q.Link.X (FKZ:16KIS0875). In addition, we were supported by the U.S. Department of Energy, Office of Science, National Quantum Information Science Research Centers, Quantum Systems Accelerator (QSA). AS and LD acknowledge JILA PFC funding (grant number PHY 1734006) and the EPSRC quantum career development grant EP/W028301/1, respectively.

\section*{Added notes} 
The algorithms developed in the present work are available at \cite{github}.

\newpage
\quad 
\newpage

\appendix
\section*{Appendix}
\section{A minimal example}\label{app:overlap}
In this section we consider the example presented in Fig.~\ref{fig:Fig1}(b-e), consisting of an operator
\begin{equation}\label{eqn:obs}
 O=\sigma_1^x\sigma_2^x+\sigma_1^z\sigma_2^z+\sigma_2^z=P_1+P_2+P_3.
\end{equation}
For this operator $O$ and a generic input state $\rho_{\rm in}$, we are interested in finding the best possible estimator (viz.~a cliques' cover), as well as the optimal allocation of the $M$ measurements. In order to do so, we assume that the three PS $P_j$ ($j=1,2,3$) have been previously probed infinitely many times, such that the values of $\tilde{\mathcal{Q}}_{ji} \to \Tr{\rho_{\rm in} P_j P_i}-\Tr{\rho_{\rm in} P_j}\Tr{\rho_{\rm in} P_i}$ are available beforehand. For the sake of clarity, we remark that this assumption was \textit{not} taken when using {\algo} [see Fig.~\ref{fig:Fig1}(e)]. This demonstrates that our approach, for a generic input state, is more likely to provide lower errors $\costF$ than any of the possible covers associated to the operator $O$ in Eq.~\eqref{eqn:obs}, despite the disadvantage of not knowing $\tilde{\mathcal{Q}}$ beforehand.

For a generic input state $\rho_{\rm in}$, one out of the four covers presented in Fig.~\ref{fig:Fig1}(b-d) resorts to the smallest error $\costF_{\rm opt}$, provided the $M$ measurements have been allocated optimally. The idea is to find this $\costF_{\rm opt}$ and compare it with the errors $\costF$ determined with the different covers in Fig.~\ref{fig:Fig1}(b-d) and {\algo} in Fig.~\ref{fig:Fig1}(e). The relative difference $[\costF-\costF_{\rm opt}] / \costF_{\rm opt}$ plotted in the histograms is thus representative of how well the associated covers (or {\algo}) perform with respect to a wide variety of random input states $\rho_{\rm in}$. Specifically, $\rho_{\rm in}$ correspond to pure states uniformly distributed on the Bloch sphere.

As a first step, we describe how to determine the optimal allocation of the $M$ measurements for any clique cover. We start by rewriting the errors $\costF$ associated to the four different covers as a sum of the elements in the matrices (recall $c_j = 1$ for all $j=1,2,3$)
\begin{align}
 \costF &= \sum
  \begin{bmatrix}
   \frac{\tilde{\mathcal{Q}}_{11}}{\xi_{1}} &
   0 & 
   0 \\
   0 &
   \frac{\tilde{\mathcal{Q}}_{22}}{\xi_{2}} & 
   0 \\
   0 &
   0 & 
   \frac{\tilde{\mathcal{Q}}_{33}}{\xi_{3}} 
   \end{bmatrix}
   \text{ for Fig.~\ref{fig:Fig1}(b)}, \label{eq:error_single_clique} \\
   \costF &=
   \begin{cases}
  \sum \begin{bmatrix}
   \frac{\tilde{\mathcal{Q}}_{11}}{\xi_{1}} &
   \frac{\tilde{\mathcal{Q}}_{12}}{\xi_{1}} & 
   0 \\
   \frac{\tilde{\mathcal{Q}}_{12}}{\xi_{1}} &
   \frac{\tilde{\mathcal{Q}}_{22}}{\xi_{1}} & 
   0 \\
   0 &
   0 & 
   \frac{\tilde{\mathcal{Q}}_{33}}{\xi_{2}} 
   \end{bmatrix} \\
   \sum \begin{bmatrix}
   \frac{\tilde{\mathcal{Q}}_{11}}{\xi_{1}} &
   0 & 
   0 \\
   0 &
   \frac{\tilde{\mathcal{Q}}_{22}}{\xi_{2}} & 
   \frac{\tilde{\mathcal{Q}}_{23}}{\xi_{2}} \\
   0 &
   \frac{\tilde{\mathcal{Q}}_{23}}{\xi_{2}} & 
   \frac{\tilde{\mathcal{Q}}_{33}}{\xi_{2}} 
   \end{bmatrix}
   \end{cases}
   \text{ for Fig.~\ref{fig:Fig1}(c)}, \label{eq:error_double_clique} \\
   \costF &= \sum
  \begin{bmatrix}
   \frac{\tilde{\mathcal{Q}}_{11}}{\xi_{1}} &
   \frac{\tilde{\mathcal{Q}}_{12}}{\xi_{1}+\xi_{2}} & 
   0 \\
   \frac{\tilde{\mathcal{Q}}_{12}}{\xi_{1}+\xi_{2}} &
   \frac{\tilde{\mathcal{Q}}_{22}}{\xi_{1}+\xi_{2}} & 
   \frac{\tilde{\mathcal{Q}}_{23}}{\xi_{1}+\xi_{2}} \\
   0 &
   \frac{\tilde{\mathcal{Q}}_{23}}{\xi_{1}+\xi_{2}} & 
   \frac{\tilde{\mathcal{Q}}_{33}}{\xi_{2}} 
   \end{bmatrix}
   \text{ for Fig.~\ref{fig:Fig1}(d)}, \label{eq:error_max_clique}
\end{align}
where the two alternatives in the curly bracket in Eq.~\eqref{eq:error_double_clique} are used to indicate the two possible cases where one clique is maximal and the other is not, as depicted in Fig.~\ref{fig:Fig1}(c). These matrices correspond to the ones pictorially represented in Fig.~\ref{fig:Fig1}(b-d), and fully determine the expected error $\costF$ for any choice of $\xi_1$, $\xi_2$ and (for the cover without maximal cliques) $\xi_3$ [see Eqs.\eqref{eq:gen_error}]. These last parameters represent the number of times that the corresponding clique is measured, and can be obtained from $m_j$ and $m_{ji}$ (and vice versa, see Sec.~\ref{app:variance_calc}). From Eqs.~\eqref{eq:error_single_clique}, \eqref{eq:error_double_clique} and \eqref{eq:error_max_clique} it is possible to determine the cliques in which the PS are by looking at the subscripts of the parameters $\xi_j$ in the diagonal elements. If $\tilde{\mathcal{Q}}_{jj}$ is divided by $\xi_1$, $\xi_2$ or $\xi_3$, it means that the PS $P_j$ belongs to the first, second or third clique, respectively. Similarly, in Eq.~\eqref{eq:error_max_clique}, the element $\tilde{\mathcal{Q}}_{22}/(\xi_1 + \xi_2)$ indicates that the PS $P_2$ is measured any time either clique is probed, in agreement with Fig.~\ref{fig:Fig1}(d).

In order to determine the optimal allocation of the $M$ measurements for any of the covers in Fig.~\ref{fig:Fig1}(b-d), we minimize the corresponding error $\costF$ in Eqs.~\eqref{eq:error_single_clique}, \eqref{eq:error_double_clique} or \eqref{eq:error_max_clique} with the additional constraints that $\sum_j \xi_j = M$ and that $\xi_j$ are non-negative integers for all $j$. We note that a similar problem (without allowing for an overlap of cliques) has been studied in Ref.~\cite{2018_RBM}. Since we are assuming that we know the exact values of $\tilde{\mathcal{Q}}_{ji}$, the errors $\costF$ derived with the optimal $\xi_j$ are the minima of the associated cover that can be achieved with the considered input state $\rho_{\rm in}$ and the measurement budget $M$. Furthermore, the four covers presented in Fig.~\ref{fig:Fig1}(b-d) are all non-redundant covers that can be found for the operator $O$ in Eq.~\eqref{eqn:obs}, meaning that out of these four $\costF$, the smallest one is the minimal error $\costF_{\rm opt}$ that can be generally obtained when measuring $O$ with $M$ shots with respect to $\rho_{\rm in}$.

Once the lowest possible errors associated to each cover have been determined, it is possible to test whether there is one cover always resorting to the minimal error $\costF_{\rm opt}$. One may think that the cover in Fig.~\ref{fig:Fig1}(d), consisting of maximal cliques only, always outperforms the others, since it probes more PS per shot. However, this is not the case due to possibly positive covariances $\tilde{\mathcal{Q}}_{ji}$ ($j\neq i$) in the off-diagonal terms in Eq.~\eqref{eq:error_max_clique}\footnote{We remark that, in general, the best measurement strategy is the one measuring \textit{all} PS at once. However, when some of the PS do not commute and as such their covariances cannot contribute to $\costF$, it is sometimes advantageous \textit{not} to measure a PS for the reasons explained in Sec.~\ref{sec:theory}.}. In fact, it turns out that the error from the cover made of maximal cliques only equals the minimal $\costF_{\rm opt}$ $\sim 54\%$ of the times. The other $\sim 46\%$ of times $\costF_{\rm opt}$ corresponds to one of the two covers in Fig.~\ref{fig:Fig1}(c).

Despite not always being the optimal one, the cover made of maximal cliques does outperform all other covers on average. This motivated the choice of using maximal cliques only when probing the system, along with the post-processing unit afterwards for removing vertices from the cliques based on their covariances and their contribution to $\costF$ within the considered clique. In fact, by comparing Fig.~\ref{fig:Fig1}(e) with Fig.~\ref{fig:Fig1}(b-d), we see that on average {\algo} provides smaller errors if compared to the theoretical minima that can be obtained with the four covers considered in the figure.

\section{Bayesian estimation for the graph}\label{app:bayes}
In this section, we use Bayesian inference \cite{kruschke2014doing,1995_GCSR} to estimate the averages, variances and covariances of the PS spanning $O$. 
We consider a pair of PS $P_j$ and $P_i$ whose measurements are collected in the dataset $D$. From $D$, it is possible to obtain the following quantities. We denote by $s_\lambda^{\pm}$ (with $s_\lambda^{+}+s_\lambda^{-} = m_\lambda$) the total number of times an outcome $\pm 1$ is collected from measurements of $P_\lambda$ for $\lambda =j,i$. The associated underlying (but unknown) probability is denoted by $\theta_\pm(\lambda)$, where the argument $\lambda$ will be dropped for clarity whenever it is clear from context. Similarly, $s_{ji}^{\pm\pm}$ (with $s_{ji}^{++}+s_{ji}^{+-}+s_{ji}^{-+}+s_{ji}^{+-} = m_{ji}$) is the number of times that $P_j$ and $P_i$ yielded the corresponding $\{\pm1,\pm1\}$ outcome, which is characterized by a probability $\theta_{\pm\pm}$. Finally, $w_\lambda^{\pm}$ is the number of occurrences of $\pm 1$, that refers to the cases where $P_j$ and $P_i$ have been probed independently. It follows that $w_j^{+}+w_j^{-} = m_j - m_{ji}$ and $w_i^{+}+w_i^{-} = m_i - m_{ji}$.

Using the collected data $D$, our goal is to estimate the posterior probability $P(\vec{\theta}\vert D)$ that best describes the parameters $$\vec{\theta} = \lbrace \theta_+(j),\theta_-(j),\theta_+(i),\theta_-(i),\theta_{++},\theta_{+-},\theta_{-+},\theta_{--} \rbrace$$ in our measurement process. By Bayes' Theorem,
\begin{equation}\label{eq:bayesian_fundamental}
    P(\vec{\theta} \vert D) = \frac{P(D \vert \vec{\theta})P(\vec{\theta})}{P(D)},
\end{equation}
where the likelihood $P(D\vert\vec{\theta})$ is the probability of obtaining the dataset $D$ given $\vec{\theta}$. The prior $P(\vec{\theta})$ is the probability of $\vec{\theta}$ before the current evidence or dataset $D$ is observed. The probability $P(D)$ is also called the marginal likelihood.

Following standard procedures \cite{kruschke2014doing,1995_GCSR}, 
we consider $P(\vec{\theta}) = \Dir(K,\vec{a})$, where $\Dir$ indicates the Dirichlet distribution of order $K$
and $\vec{a} = \lbrace a_1,\dots,a_K \rbrace$ are its hyperparameters  \cite{kruschke2014doing,1995_GCSR}. Then $P(\vec{\theta}\vert D) = \Dir(K,\vec{c}+\vec{a})$, where $\vec{c} = \lbrace c_1,\ldots,c_K \rbrace$ is a vector giving the number of occurrences of each category in the dataset $D$. 

The Dirichlet distribution of order $K$ has the following probability density function:
\begin{subequations}
\begin{align}
P(\vec{\theta}) & = P(\vec{\theta};\vec{a}) = \frac{1}{B(\vec{a})}\prod_{i=1}^K\theta_i^{a_i-1}, 
\label{eqn:prior}\\
B(\vec{a}) & = \frac{\prod_{i=1}^K\Gamma(a_i)}{\Gamma(\sum_{i=1}^K a_i)},
\label{eqn:B}
\end{align}
\end{subequations}
where $\Gamma(z) = \int_{0}^{\infty} x^{z-1} e^{-x} dx $ is the gamma function. The likelyhood function $P(D\vert\vec{\theta})$ then takes the form
\begin{equation}\label{eqn:likelihood}
P(D\vert\vec{\theta})=\prod_{i=1}^K\theta_i^{c_i},
\end{equation}
which leads to the posterior
\begin{equation}\label{eqn:posterior}
P(\vec{\theta}\vert D)=\frac{1}{B(\vec{a}+\vec{c})}\prod_{i=1}^K\theta_i^{c_i+a_i-1}.
\end{equation}

With the posterior probability $P(\vec{\theta}\vert D)$ in Eq.~\eqref{eqn:posterior}, we can determine the quantities of interest for the measurement protocol considered in this work by marginalization. In more detail, assume we are interested in estimating a certain quantity $f(\vec{\theta})$. We define its estimator $\tilde{f}(\vec{\theta})$ by
\begin{equation}\label{eq:gen_estimator_bayes}
    \tilde{f}(\vec{\theta}) = \int f(\vec{\theta}) P(\vec{\theta}\vert D) d \vec{\theta}.
\end{equation}
This equation is used for determining $\tilde{P_j}$ and the elements of the covariance matrix $\tilde{\mathcal{Q}}$ (see Sec.~\ref{sec:theory}). 

In the following, we explicitly derive $\tilde{P_j}$ and the variance $\tilde{\mathcal{Q}}_{jj}$ of a single PS $P_j$ in Sec.~\ref{app:sec:single}. 
In Sec.~\ref{app:sec:two_together}, we consider the case in which two PS $P_j$ and $P_i$ are measured together, and we estimate their covariance.
As explained in Sec.~\ref{app:sec:two_general}, this estimate for their covariance may lead to wrong errors $\costF$ when $M$ is small. Then, we obtain a valid estimate for $\tilde{\mathcal{Q}}_{ji}$ ($j \neq i$) by considering the case in which the PS may also have been measured independently from each other. Finally, we summarize the results and explain how Bayesian estimation is used in {\algo} in Sec.~\ref{app:sec:our_algo_features}.

\subsection{Single Pauli string}\label{app:sec:single}
Whenever we measure a single PS $P_j$, we obtain a dataset $D$ which is a collection of $+1$ and $-1$, from which $s_j^+$ and $s_j^-$ can be found. In this case, $K=2$, $\vec{\theta}=\lbrace \theta_+, \theta_-\rbrace$ with $\theta_-=1-\theta_+$, and for simplicity we set $D \mapsto \vec{c}=\lbrace s_j^+, s_j^- \rbrace$. From Eq.~\eqref{eqn:posterior} we get the posterior probability
\begin{equation}\label{eq:bayes_single_gen}
\begin{split}
    P&(\lbrace \theta_+,\theta_- \rbrace \vert \lbrace s_j^+,s_j^- \rbrace)   = \frac{\theta_+^{s_j^+}(1-\theta_+)^{s_j^-}}{B(\lbrace s_j^+ + 1,s_j^- +1 \rbrace)},
\end{split}
\end{equation}
where we set $a_1=a_2=1$. This choice for $\vec{a}$, corresponding to a uniformly distributed prior, is generally taken when no information is available prior to the measurement. 

The exact values $\langle P_j \rangle$ and $(\Delta P_j)^2$ of the average and variance of the PS $P_j$ can be expressed in terms of the probabilities $\vec{\theta}$ as
\begin{subequations}\label{eq:bayes_avg_var_theta}
\begin{align}
    \langle P_j \rangle & = \theta_+ - \theta_- = 2 \theta_+ - 1, \label{eq:bayes_avg_var_theta_avg} \\ 
    (\Delta P_j)^2 & = 4 \theta_+ \theta_- = 4 \theta_+ (1 - \theta_+) .\label{eq:bayes_avg_var_theta_var}
\end{align}
\end{subequations}
Using these equations in place of $f(\vec{\theta})$ in Eq.~\eqref{eq:gen_estimator_bayes}, we obtain the following estimates $\tilde{P}_j$ and $\tilde{\mathcal{Q}}_{jj}$:
\begin{subequations}\label{eq:estimators_single_Pauli}
\begin{align}
\begin{split}
    \tilde{P}_j & = \int_{0}^{1} \langle P_j \rangle P(\lbrace \theta_+,\theta_- \rbrace \vert \lbrace s_j^+,s_j^- \rbrace) d \theta_+ \\ & = \frac{s_j^+ - s_j^-}{s_j^+ + s_j^- + 2} ,
\end{split}\\ 
\begin{split}
    \tilde{\mathcal{Q}}_{jj} & = \int_{0}^{1} (\Delta P_j)^2 P(\lbrace \theta_+,\theta_- \rbrace \vert \lbrace s_j^+,s_j^- \rbrace) d \theta_+ \\ & = 4\frac{\left( s_j^+ + 1 \right)\left( s_j^- + 1 \right)}{\left( s_j^+ + s_j^- + 2 \right)\left( s_j^+ + s_j^- + 3 \right)}. \label{eq:estimators_single_Pauli_var}
\end{split}
\end{align}
\end{subequations}
These relations are used by our algorithm to update vertices and self-edges in the graph [see, e.g., Fig.~\ref{fig:Fig1}(a)].

\subsection{Two PS always measured together}\label{app:sec:two_together}
The procedure used in the previous section to determine $\tilde{P}_j$ and $\tilde{\mathcal{Q}}_{jj}$ can be repeated here for the covariance $\tilde{\mathcal{Q}}_{ji}$ ($j\neq i$), provided we have meaningful likelihood and prior functions. In this case, we deal with two PS $P_j$ and $P_i$, and as such we have four possible outcomes described by the probabilities $\theta_{++}$, $\theta_{+-}$, $\theta_{-+}$ and $\theta_{--} = 1-\theta_{++}-\theta_{+-}-\theta_{-+} $. Furthermore, we assume here that $P_j$ and $P_i$ are always measured together; the generalization is presented in Sec.~\ref{app:sec:two_general}. For the sake of clarity, we have $K=4$, $\vec{\theta} = \lbrace \theta_{++},\theta_{+-},\theta_{-+},\theta_{--}\rbrace$ and $D \mapsto \vec{c} = \lbrace s_{ji}^{++},s_{ji}^{+-},s_{ji}^{-+},s_{ji}^{--} \rbrace$. From Eq.~\eqref{eqn:posterior} we have the following posterior probability,
\begin{equation}\label{eq:bayes_together_gen}
\begin{split}
    P&(\vec{\theta}\vert D) = \theta_{++}^{s_{ji}^{++}+a_{++}-1}\theta_{+-}^{s_{ji}^{+-}+a_{+-}-1}\theta_{-+}^{s_{ji}^{-+}+a_{-+}-1}  \\
    & \times (1-\theta_{++}-\theta_{+-}-\theta_{-+})^{s_{ji}^{--}+a_{--}-1} \frac{1}{B(\vec{a}+\vec{c})}  ,
\end{split}
\end{equation}
where we left the parameters $\vec{a} = \lbrace a_{++}, a_{+-}, a_{-+}, a_{--} \rbrace$ in their explicit form for reasons that will become clear in the following Sec.~\ref{app:sec:two_general}. Here, we assume $a_{++}=a_{+-}=a_{-+}=a_{--}=1$ similar to the single PS, since there is no information available prior the measurements. 

By expressing the exact covariance $\cov (P_j,P_i)$ in terms of the probabilities $\vec{\theta}$ as
\begin{equation}\label{eq:exact_cov_theta}
\begin{split}
    \cov & (P_j,P_i) = 4( \theta_{++}\theta_{--} - \theta_{+-}\theta_{-+}) \\ 
    & = 4\left[ \theta_{++}(1-\theta_{++}-\theta_{+-}-\theta_{-+}) - \theta_{+-}\theta_{-+}\right],
\end{split}
\end{equation}
and using Eqs.~\eqref{eq:gen_estimator_bayes} and \eqref{eq:bayes_together_gen}, we can determine a possible estimate $\tilde{\mathcal{Q}}_{ji}$ ($j\neq i$),
\begin{widetext}
\begin{align}\label{eq:wrong_estimate_cov}
\begin{aligned}
    \tilde{\mathcal{Q}}_{ji} &= \int_0^1 d\theta_{++} \int_0^{1-\theta_{++}} d\theta_{+-} \int_0^{1-\theta_{++}-\theta_{+-}} d \theta_{-+}\,
    \cov (P_j,P_i) P(\vec{\theta}\vert D) \\
   &= 4\frac{(s_{ji}^{++}+1)(s_{ji}^{--}+1) - (s_{ji}^{+-}+1)(s_{ji}^{-+}+1)}{(s_{ji}^{++} + s_{ji}^{+-} + s_{ji}^{-+} + s_{ji}^{--} + 4)(s_{ji}^{++} + s_{ji}^{+-} + s_{ji}^{-+} + s_{ji}^{--} + 5)}.
\end{aligned}
\end{align}
\end{widetext}

This equation, while representing a valid estimate for the covariance, is \emph{not} used in our algorithm to obtain $\tilde{\mathcal{Q}}_{ji}$.
Indeed, since we allow for overlap among the cliques, there are cases in which two PS are both measured individually and together, and Eq.~\eqref{eq:wrong_estimate_cov} disregards all outcomes in which $P_j$ and $P_i$ are uncorrelated. In the following section, we study two possibilities on how to include these events into the estimation of $\tilde{\mathcal{Q}}_{ji}$, such that we do not neglect any collected information on the system and always obtain a meaningful estimate of the error $\costF$.

\subsection{Two PS measured together and individually}\label{app:sec:two_general}
As anticipated at the end of the previous section, Eq.~\eqref{eq:wrong_estimate_cov} ignores all events where $P_j$ and $P_i$ were not measured together. This has two implications. First, we are discarding available information that can be used for getting a more precise estimate for the covariance. Second, and more importantly, in case of scarce statistics Eqs.~\eqref{eq:wrong_estimate_cov} and \eqref{eq:estimators_single_Pauli} may lead to a wrong estimate for the error $\costF$.
This is because the variances $\tilde{\mathcal{Q}}_{jj}$ and the covariances $\tilde{\mathcal{Q}}_{ji}$ ($j\neq i$) are derived independently from each other, ignoring the fact that they are correlated. In fact, it is possible to express the probabilities $\theta_{\pm\pm}$ in terms of $\theta_{\pm}(j)$, $\theta_{\pm}(i)$, and the exact $\cov (P_j,P_i)$:
\begin{subequations}\label{eq:theta_relations}
\begin{align}
    \theta_{++} & = \theta_{+}(j)\theta_{+}(i) + \frac{\cov (P_j,P_i)}{4},\\
    \theta_{+-} & = \theta_{+}(j)\theta_{-}(i) - \frac{\cov (P_j,P_i)}{4},\\
    \theta_{-+} & = \theta_{-}(j)\theta_{+}(i) - \frac{\cov (P_j,P_i)}{4},\\
    \theta_{--} & = \theta_{-}(j)\theta_{-}(i) + \frac{\cov (P_j,P_i)}{4},
\end{align}
\end{subequations}
where we recall that the subscripts $\pm$ in $\theta_{\pm\pm}$ refer to PS $P_j$ and $P_i$, in this order. Since the relations in Eqs.~\eqref{eq:theta_relations} are not always satisfied by Eqs.~\eqref{eq:wrong_estimate_cov} and \eqref{eq:estimators_single_Pauli}, it may happen that in measuring an observable we obtain an estimate resulting in a wrong error $\costF$. As an example, assume that we take $25$ measurements of each of $P_1$ and $P_2$ independently, and always get $+1$ and $-1$ outcomes, respectively. Afterwards, we measure these PS together twice and get the two pairs $\lbrace -1,+1 \rbrace$ and $\lbrace +1,-1 \rbrace$. In this case, our estimates for the variances will be small, while we would get a comparatively large negative covariance. This leads to a negatively defined covariance matrix $\tilde{\mathcal{Q}}$ such that $\tilde{\mathcal{Q}}_{11}+\tilde{\mathcal{Q}}_{22}+2\tilde{\mathcal{Q}}_{12} = -0.07$, which is not physically allowed.

To prevent this from happening, we describe two alternative ways for estimating the covariance. The first one is more precise but unpractical, since it yields a result that cannot be efficiently evaluated numerically. The second is less precise, but computationally advantageous.

The first way consists of defining a probability distribution function describing the measurements of two PS both independently and simultaneously probed, in which Eqs.~\eqref{eq:theta_relations} are satisfied. To do so, we start from the joint posterior distribution $P(\vec{\theta}\vert D)$:
\begin{equation}\label{eq:bayes_toghether_generalprior}
\begin{split}
    P & (\vec{\theta}\vert D) = \frac{1}{B(\vec{a}+\vec{c})} \Big[ \\
    & \theta_{++}^{s_{ji}^{++}+a_{++}-1}\theta_{+-}^{s_{ji}^{+-}+a_{+-}-1}\theta_{-+}^{s_{ji}^{-+}+a_{-+}-1} \theta_{--}^{s_{ji}^{--}+a_{--}-1} \\
    &\times \theta_{+}(j)^{w_j^+ +a_{j+}-1} \theta_{-}(j)^{w_j^- +a_{j-}-1} \\
    &\times \theta_{+}(i)^{w_i^+ +a_{i+}-1} \theta_{-}(i)^{w_i^- +a_{i-}-1} \Big],
\end{split}
\end{equation}
where the order is $K=8$, and
\begin{subequations}
\begin{align}
    D \mapsto \vec{c} & = \lbrace s_{ji}^{++},s_{ji}^{+-},s_{ji}^{-+},s_{ji}^{--},w_i^+,w_i^-,w_j^+,w_j^- \rbrace, \nonumber \\
    \vec{\theta} & = \lbrace\theta_{++},\theta_{+-},\theta_{-+},\theta_{--},\theta_{i+},\theta_{i-},\theta_{j+},\theta_{j-} \rbrace, \nonumber \\
    \vec{a} & = \lbrace a_{++},a_{+-},a_{-+},a_{--},a_{j+},a_{j-},a_{i+},a_{i-} \rbrace . \nonumber
\end{align}
\end{subequations}
Here, $\theta_{--}=1-\theta_{++}-\theta_{+-}-\theta_{-+}$, $\theta_{-}(\lambda)=1-\theta_{+}(\lambda)$ for $\lambda = j,i$, and we set $\vec{a}$ to be the all ones vector in the following. Evidently, $P(\vec{\theta}\vert D)$ in Eq.~\eqref{eq:bayes_toghether_generalprior} is the product of the three distinct distributions associated to $P_j$ and $P_i$ being measured together (second row), and independently (third and fourth rows). In particular, since $K=8$, there are eight probabilities in $\vec{\theta}$, but only four are independent. Indeed, by inverting the relations in Eqs.~\eqref{eq:theta_relations}, we obtain 
\begin{align*}
    \theta_+(j) &= \theta_{++} + \theta_{+-} &
    \theta_-(j) &= \theta_{--} + \theta_{-+}\\
    \theta_+(i) &= \theta_{++} + \theta_{-+} &
    \theta_-(i) &= \theta_{--} + \theta_{+-}.
\end{align*} 
Using these, we can define a new probability distribution $\mathcal{P}(\vec{\theta}\vert D)$ in the following way,
\begin{subequations} \label{eq:bayes_two_together}
\begin{align}
\begin{split}\label{eq:bayes_two_together_likelyhood} 
    \mathcal{P}(\vec{\theta}\vert D) = & \frac{1}{\mathcal{N}} \Big[
    \theta_{++}^{s_{ji}^{++}}\theta_{+-}^{s_{ji}^{+-}}\theta_{-+}^{s_{ji}^{-+}} \theta_{--}^{s_{ji}^{--}} \\
    &\times (\theta_{++} + \theta_{+-})^{w_j^+} (\theta_{--} + \theta_{-+})^{w_j^-} \\
    &\times (\theta_{++} + \theta_{-+})^{w_i^+} (\theta_{--} + \theta_{+-})^{w_i^-} \Big],
\end{split}\\
\begin{split}\label{eq:bayes_two_together_norm}
    \mathcal{N} = \int_{0}^{1} d  & \theta_{++} \int_{0}^{1-\theta_{++}} d\theta_{+-} \int_{0}^{1-\theta_{++}-\theta_{+-}} d\theta_{-+}  \\ \times \theta_{++}^{s_{ji}^{++}} & \theta_{+-}^{s_{ji}^{+-}}\theta_{-+}^{s_{ji}^{-+}}(1-\theta_{++}-\theta_{+-}-\theta_{-+})^{s_{ji}^{--}}.
\end{split}
\end{align}
\end{subequations}
Here, only the independent variables $\theta_{\pm\pm}$ are used to describe the measurements.

From $\mathcal{P}(\vec{\theta}\vert D)$, repeating the same steps as in Sec.~\ref{app:sec:two_together}, it is then possible to determine the analytical form of the expected covariance $\tilde{\mathcal{Q}}_{ji}$ ($j \neq i$), when including events in which the PS $P_j$ and $P_i$ are both measured independently and together. This form of the covariance (which is lengthy and thus not explicitly reported) ensures that the error $\costF$ and the covariance matrix $\tilde{\mathcal{Q}}$ are \textit{always} well defined and rapidly converge to the exact values. However, since the probability distribution function $\mathcal{P}(\vec{\theta}\vert D)$ in Eq.~\eqref{eq:bayes_two_together} is not a Dirichlet distribution, the resulting form of the covariance cannot be efficiently computed numerically. In more details, we obtain a complicated sum of factorials which we were unable to simplify, and that requires a long time to be evaluated when the dataset $D$ is large.

There is a second alternative to estimate the covariance that both includes the information from the PS being measured independently, and avoids getting wrong errors $\costF$. Rather than defining a joint probability distribution as before, we modify the hyperparameters $\vec{a}$ in the prior in Eq.~\eqref{eqn:prior} to obtain a more precise posterior distribution $P(\vec{\theta}\vert D)$. In other words, we are going to modify the values of $\vec{a}$ in Eq.~\eqref{eq:bayes_together_gen} depending on the outcomes $w_j^\pm$ and $w_i^\pm$. As such, the dataset described by $\vec{c}$, as well as the probabilities $\vec{\theta}$, are going to be the same as in Sec.~\ref{app:sec:two_together}.

The first step consists in observing that, in the absence of measurements, $\vec{c} = \vec{0} = \lbrace 0, 0, 0, 0 \rbrace$, the estimates for the probabilities $\vec{\theta}$ are [using the posterior in Eq.~\eqref{eq:bayes_together_gen}]
%
\begin{equation}\label{eq:prior_theta_exp}
    \tilde{\theta}_{\pm\mp} = \frac{a_{\pm\mp}}{a_{++}+a_{+-}+a_{-+}+a_{--}} \text{ for $\vec{c} = \vec{0}$},
\end{equation}
which are all equal to $1/4$ for $\vec{a} = \lbrace 1,1,1,1 \rbrace$ (as expected when  there is no information about the measurements). However, in this case, we do have information available, coming from all the times in which the PS $P_j$ and $P_i$ have been independently probed. In particular, by using the procedure outlined in Sec.~\ref{app:sec:single} with the substitution $s_{\lambda}^{\pm}\mapsto w_{\lambda}^{\pm}$, we can determine the expected values $\tilde{\theta}_{\pm}(\lambda)$ for $\theta_{\pm}(\lambda)$ ($\lambda = j,i$) as follows\footnote{Notice that, to estimate $\theta_{\pm}(j)$, we only use the events in which the PS $P_j$ has been measured independently from $P_i$ (and vice versa). This is done to avoid biases that could falsify the error estimate $\costF$.},
\begin{subequations} \label{eq:pfrom_exp_theta_exp-1}
\begin{align}
   \tilde{\theta}_{\pm}(j) = & \frac{w_j^\pm + 1}{w_j^+ + w_j^- +2} ,\\
   \tilde{\theta}_{\pm}(i) = & \frac{w_i^\pm + 1}{w_i^+ + w_i^- +2} .
\end{align}
\end{subequations}
The idea is now to plug these $\tilde{\theta}_{\pm}(\lambda)$ into Eqs.~\eqref{eq:theta_relations}, determine the expected values for $\theta_{\pm\pm}$, and then use those to find the best hyperparameters $\vec{a}$ from Eqs.~\eqref{eq:prior_theta_exp}. To do so, however, we need two more ingredients. First, as it is possible to see from Eqs.~\eqref{eq:theta_relations}, we also require an estimate for the exact covariance $\cov (P_j,P_i)$. Since independent measurements of $P_j$ and $P_i$ are uncorrelated, we set $\cov (P_j,P_i)$ to be zero as our initial guess. Second, we are free to choose a normalization $A = a_{++} + a_{+-} + a_{-+} + a_{--}$ that fixes the bias towards the prior in determining the estimates from the resulting posterior distribution \cite{kruschke2014doing,1995_GCSR}. Since we want to calculate $\tilde{\mathcal{Q}}_{ji}$ ($j \neq i$), but our prior assumes zero covariance, we set $A=4$ in the following, which is the same normalization generally used when no knowledge is available (in fact, we set $A=K$ throughout this work).

From the above considerations, we determine the elements of $\vec{a}$ to be
%
\begin{equation}\label{eq:pfrom_exp_theta_exp}
    a_{\pm\mp} = 4\frac{\left(1+w_{j}^\pm\right)\left(1+w_{i}^\mp\right)}{\left(2+w_{j}^++w_{j}^-\right)\left(2+w_{i}^++w_{i}^-\right)} .
\end{equation}
By using these values of $\vec{a}$ in Eq.~\eqref{eq:bayes_together_gen} we can obtain the posterior distribution $P(\vec{\theta}\vert D) $, from which it is possible to find [see Eq.~\eqref{eq:gen_estimator_bayes} or the first row in Eq.~\eqref{eq:wrong_estimate_cov}]
\begin{widetext}
\begin{align}\label{eq:used_estimate_cov}
\begin{aligned}
    \tilde{\mathcal{Q}}_{ji} &= 4\frac{(s_{ji}^{++}+a_{++})(s_{ji}^{--}+a_{--}) - (s_{ji}^{+-}+a_{+-})(s_{ji}^{-+}+a_{-+})}{(s_{ji}^{++} + s_{ji}^{+-} + s_{ji}^{-+} + s_{ji}^{--} + 4)(s_{ji}^{++} + s_{ji}^{+-} + s_{ji}^{-+} + s_{ji}^{--} + 5)}.
\end{aligned}
\end{align}
\end{widetext}
Compared to the estimate for $\tilde{\mathcal{Q}}_{ji}$ given in Eq.~\eqref{eq:wrong_estimate_cov}, this equation provides more accurate results, particularly when the dataset $D$ is small. Furthermore, it lowers the probability of getting nonphysical values for the covariance matrix $\tilde{\mathcal{Q}}$ and the cost function $\costF$. However, since the relations in Eqs.~\eqref{eq:theta_relations} are still not directly enforced into the posterior probability $P(\vec{\theta}\vert D)$, there are still cases in which Nature conspires against us to get a nonphysical $\tilde{\mathcal{Q}}$. In addition to that, the post-processing subroutine in our algorithm (see Fig.~\ref{fig:Fig2} and Sec.~\ref{sec:algorithm}) is particularly efficient in finding these cases, since it tries to minimize $\costF$. 

To make the wrong estimation of the covariances statistically impossible, we assign a risk factor $R_{ji}$ for each pair of PS $P_j$ and $P_i$, proportional to how likely the corresponding covariance matrix $\tilde{\mathcal{Q}}$ is to be nonphysical. We define $R_{ji}$ by
\begin{subequations} \label{eq:risk_def}
\begin{align}
   R_{ji} &= \text{sign}(c_j c_i)\beta_1\left[ 1- \frac{1}{1+e^{-\beta_2(r_{ji}-\beta_3)}} \right] ,\\
   r_{ji} &= \frac{s_{ji}^{++} + s_{ji}^{+-} + s_{ji}^{-+} + s_{ji}^{--}}{w_j^{+} + w_j^{-} + w_i^{+} + w_i^{-}} ,
\end{align}
\end{subequations}
where $\text{sign}(x)$ is the sign of $x$, and $\beta_1$, $\beta_2$ and $\beta_3$ are arbitrary parameters \footnote{We found that by setting $\beta_1$, $\beta_2$ and $\beta_3$ to $1$, $10$ and $0.75$, respectively, the probability of obtaining unphysical covariances is practically zero.}. This definition of $R_{ji}$ is such that it assigns a higher risk (in absolute value) when $P_j$ and $P_i$ have been measured more times individually than together. Once $R_{ji}$ is assigned to the pair $P_j$ and $P_i$, we increase $\tilde{\mathcal{Q}}_{ji}$ by $R_{ji}$ times 
$\delta\tilde{\mathcal{Q}}_{ji} = \sqrt{(\tilde{\mathcal{Q}}^2)_{ji}-(\tilde{\mathcal{Q}}_{ji})^2}$, where $(\tilde{\mathcal{Q}}^2)_{ji}$ is the estimated square of the covariance $\cov^2(P_j,P_i)$:
\begin{equation}\label{eq:covS}
\begin{split}
    \cov^2(P_j,P_i) & = \expval{\left(P_jP_i-\expval{P_j}\expval{P_i}\right)^2} \\ & = 16(\theta_{++}\theta_{--} - \theta_{+-}\theta_{-+})^2.
\end{split}
\end{equation}
$(\tilde{\mathcal{Q}}^2)_{ji}$ can be found from Eq.~\eqref{eq:bayes_together_gen} by using the usual approach presented in Eq.~\eqref{eq:gen_estimator_bayes}, and is equal to
\begin{widetext}
\begin{subequations}\label{eq:covsq}
\small
\begin{align}
   (\tilde{\mathcal{Q}}^2)_{ji} & = \frac{\text{Num}}{\text{Den}} ,\\
   \begin{split}
    \text{Num} & = 16 \left\lbrace (a_4 + 
    s_{ji}^{--}) (1 + a_4 + s_{ji}^{--}) (s_{ji}^{++})^2  + (a_3 + s_{ji}^{-+}) (1 + a_3 + s_{ji}^{-+}) s_{ji}^{+-} (1 + 
    s_{ji}^{+-}) \right. \\
    & \left. + a_2 (a_3 + s_{ji}^{-+}) (1 + a_3 + s_{ji}^{-+} + 2 s_{ji}^{+-} + 2 (a_3 + s_{ji}^{-+}) s_{ji}^{+-} - 
    2 (a_4 + s_{ji}^{--}) (a_1 + s_{ji}^{++})) \right. \\
    & \left. + (a_4 + s_{ji}^{--}) (1 + a_4 + s_{ji}^{--} - 2 (a_3 + s_{ji}^{-+}) s_{ji}^{+-}) s_{ji}^{++}  + 
    a_2^2 (a_3 + s_{ji}^{-+}) (1 + a_3 + s_{ji}^{-+}) \right. \\
    & \left. + 
    a_1 (a_4 + s_{ji}^{--}) (1 + a_4 + s_{ji}^{--} - 2 (a_3 + s_{ji}^{-+}) s_{ji}^{+-} + 2 s_{ji}^{++} + 
    2 (a_4 + s_{ji}^{--}) s_{ji}^{++}) \right. \\
    & \left.  + a_1^2 (a_4 + s_{ji}^{--}) (1 + a_4 + s_{ji}^{--})   \right\rbrace,
    \end{split} \\
    \begin{split}
        \text{Den} & = (a_1 + a_2 + a_3 + a_4 + s_{ji}^{++} + s_{ji}^{+-} + s_{ji}^{-+} + s_{ji}^{--})(a_1 + a_2 + a_3 + a_4 + s_{ji}^{++} + s_{ji}^{+-} + s_{ji}^{-+} + s_{ji}^{--}+1) \\
        & \times (a_1 + a_2 + a_3 + a_4 + s_{ji}^{++} + s_{ji}^{+-} + s_{ji}^{-+} + s_{ji}^{--}+2)(a_1 + a_2 + a_3 + a_4 + s_{ji}^{++} + s_{ji}^{+-} + s_{ji}^{-+} + s_{ji}^{--}+3) ,
    \end{split}
\end{align}
\end{subequations}
\end{widetext}
where the elements of $\vec{a}$ are defined in Eqs.~\eqref{eq:pfrom_exp_theta_exp}. Since $\delta\tilde{\mathcal{Q}}_{ji}$ is a statistical error on the covariance $\tilde{\mathcal{Q}}_{ji}$, it approaches zero when increasing the size of the dataset. 

\subsection{Summary}\label{app:sec:our_algo_features}
To summarize, we resorted to Bayesian theory to estimate averages, variances and covariances of PS being measured together and/or individually. This is crucial for allocating the measurement budget to different cliques, since our estimators yield valid results with scarce or even no statistics.

As detailed in \cite{github}, our algorithm allows for choosing between different features. Every time the covariance matrix is updated, the estimator for the covariance of two PS $P_j$ and $P_i$ is
\begin{equation}\label{eq:cov_used}
    \tilde{\mathcal{Q}}_{ji} + R_{ji} \delta\tilde{\mathcal{Q}}_{ji},
\end{equation}
where $\tilde{\mathcal{Q}}_{ji}$, $R_{ji}$ and $\delta\tilde{\mathcal{Q}}_{ji}$ are given in Eqs.~\eqref{eq:used_estimate_cov}, \eqref{eq:risk_def} and \eqref{eq:covsq}, respectively. The user has then the freedom of choosing the different hyperparameters $\beta_1$, $\beta_2$ and $\beta_3$ for the risk factor $R_{ji}$. This has implications in the allocation of the measurement budget. In our experience, setting the risk factor to zero (i.e., $\beta_1 = 0$) never resorted to unphysical covariance matrices and yielded slightly better results. Otherwise, an empirically reasonable choice for these hyperparameters is given by $\beta_1 = 1$, $\beta_2 = 10$ and $\beta_3 = 0.25$. 

Another feature of our algorithm is that, \textit{after} all measurements have been allocated and taken, it offers different options for calculating the estimator $\tilde{O}$ and its error $\costF$. The user can substitute the Bayesian estimators for $\tilde{P}_j$ and $\tilde{\mathcal{Q}}_{jj}$ in Eqs.~\eqref{eq:estimators_single_Pauli}, and for $\tilde{\mathcal{Q}}_{ji}$ ($j \neq i$) in Eq.~\eqref{eq:cov_used}, with the sample averages, variance and covariance \cite{fornasini2008uncertainty}, respectively. This is a reasonable choice since these sample quantities are known to converge faster if compared to the Bayesian ones. We remark, however, that this substitution is reasonable if and only if there is a sufficiently large dataset for correctly estimating all contributions in Eq.~\eqref{eq:gen_error_cov}.

Yet another feature of our algorithm is that it allows, both when allocating the measurement budget and/or when outputting the final value for estimate and error, to use the exact values $\cov(P_j,P_i)$ for $\tilde{\mathcal{Q}}_{ji}$ in Eq.~\eqref{eq:gen_error_cov}. On the one hand, this allows for investigating the limits of our protocol. On the other hand, it is useful to determine the exact error (i.e., without statistical fluctuations) that is expected from a given measurement allocation, as we did for the values in squared brackets in Fig.~\ref{fig:Fig3}(a) and the points in Fig.~\ref{fig:Fig3}(c).

\section{A computationally efficient method to compute $\costF$}
\label{app:variance_calc}

In this section, we explain a computationally efficient method to derive the error $\costF$. As evident from Eq.~\eqref{eq:gen_error}, this involves two objects, the matrix $\mathcal{C}$ (with elements $c_j c_i\tilde{\mathcal{Q}}_{ji}$) and the parameters $m_{ji}$ ($j,i=1,\dots,n$). The first one can be obtained by following the procedure explained in Sec.~\ref{app:bayes}. The latter one is found from the outputs of either the BF or the ML subroutine, as explained in Sec.~\ref{sec:algorithm}. Indeed, these subroutines provide an $r$-dimensional vector $\vec{\xi}$ whose elements $\xi_i$ represent the number of times that the $i$-th clique is measured. To derive $m_{ji}$ from $\vec{\xi}$, one may use the measurement characteristic matrix $\mathcal{E}$, which is an $(n\times r)$-matrix whose elements $\mathcal{E}_{ji}$ are equal to $1$ if the Pauli string $j$ belongs to the clique $i$, and zero otherwise. It then follows that $m_{ji}$ is the $(j,i)$-element of the product $\mathcal{E}\Xi\mathcal{E}^\top$, where $\Xi$ is a diagonal $(r\times r)$-matrix with $\vec{\xi}$ as the diagonal entries, and $^\top$ indicates transposition.

Given $m_{ji}$, the error $\costF$ can be rewritten as a function of matrix operations that can be efficiently implemented numerically. Denoting the Hadamard product and Hadamard division \cite{horn2012matrix} with $\HadProd$ and $\HadDiv$, respectively, we have
\begin{equation}\label{eq:loss_function}
\costF = j^\top ( \mathcal{C} \HadProd \mathcal{E}\Xi\mathcal{E}^\top \HadDiv \mathcal{E}\Xi J\Xi\mathcal{E}^\top )j  , 
\end{equation}
where $J$ is the all-ones matrix of size $(r \times r)$, and $j$ is the $(r \times 1)$-all-ones vector. Given $\mathcal{C}$ and $\vec{\xi}$, this equation is used in the BF and the ML subroutines of our algorithm to efficiently compute the error function. Notice that, compared to Eq.~\eqref{eq:gen_error}, the Kronecker delta does not appear in Eq.~\eqref{eq:loss_function}. This is not problematic since the BF [ML] subroutine considers the elements in $\vec{\xi}$ to be positive integers [real numbers that are then rounded to the nearest integer]. For the post-processing we directly use Eq.~\eqref{eq:gen_error} to calculate $\costF$.

\section{Numerical results}\label{app:our_algorithm}
In this section, we explain the numerical results in Fig.~\ref{fig:Fig3} and \ref{fig:Fig4}. This includes details on the LDF algorithm that we have used, and an extended explanation about how the values in the tables, as well as the points in the plots, have been derived. 

The numbers outside the square brackets in Fig.~\ref{fig:Fig3}(a) represent the standard deviations
\begin{equation}\label{eq:values_in_table}
    \Sigma \equiv \sqrt{\frac{\sum_{j=1}^R(\tilde{O}_j - \langle O \rangle)^2}{R}},
\end{equation}
where $\tilde{O}_j$ is the $j$-th estimated average of the observable $O$ under consideration, $\langle O \rangle$ is the exact average of $O$, and $R$ is the total number of times the whole measurement procedure is repeated [$R=40$ in Fig.~\ref{fig:Fig3}(a)]. For the Derand \cite{Huang2021Efficient} and the APS \cite{hadfield2020measurements,hadfield2021adaptive} methods, $\tilde{O}_j$ are determined with the algorithms reported in the associated references. The same holds for the OGM \cite{Bujiao2021} method, where we resorted to version $2$ of their algorithm and followed their sampling procedure.  

As explained in Sec.~\ref{app:sec:our_algo_features} and in \cite{github}, one can choose different quantities to be returned by {\algo}. The $\tilde{O}_j$ in Eq.~\eqref{eq:values_in_table} for calculating the numbers outside the square brackets are obtained with the sample average \cite{fornasini2008uncertainty} for estimating all PS $\tilde{P}_j$ within $\tilde{O}_j$. The numbers inside the square brackets in Fig.~\ref{fig:Fig3}(a), on the other side, are derived by averaging the values $\sqrt{\costF}$ obtained in the $R$ iterations of the algorithm. The associated errors in the parentheses are the sample root mean squares \cite{fornasini2008uncertainty}. While {\algo} performs the measurement allocation without prior knowledge of the covariance matrix $\tilde{\mathcal{Q}}$, the values for $\sqrt{\costF}$ that are returned at each $j$-th iteration are calculated with the exact $\tilde{\mathcal{Q}}_{j,i} \rightarrow \cov(P_j,P_i)$. This allows to have a very precise estimate of the real error (hence the small uncertainties in the parentheses), even with only $R=40$. 

In panels (b) and (c) of Fig.~\ref{fig:Fig3} we show the rescaled errors $M\Sigma^2/\langle O \rangle^2$ and $M \costF / \langle O \rangle^2$, respectively. In (b), $M\Sigma^2/\langle O \rangle^2$ is chosen in order to compare {\algo} with other approaches that cannot directly estimate the variance of the considered observable $O$. In (c), we report $M \costF / \langle O \rangle^2$, where $\costF$ and the associated error bars are calculated in the same way as the values in the square brackets within Fig.~\ref{fig:Fig3}(a) (see previous paragraph). We have chosen $R=100$ to reduce statistical fluctuations affecting $\Sigma$ in (b), and $R=40$ in (c). Furthermore, since the APS method and {\algo} with the BF subroutine allocate the measurement shots one at the time, when $M$ is large they require longer runtimes than the other approaches. Therefore, they have been run up to $M=6400$. Every time {\algo} is used to determine the $\tilde{\mathcal{Q}}$ used for measurement allocation, it resorts to Bayesian estimation with $\beta_1 = 0$ (see Sec.~\ref{app:sec:our_algo_features}).

For observables characterized by small values of $n$, {\algo} can find all maximal cliques and feed them into either the BF or the ML subroutines for measurement allocation. This procedure led to the reported values in Fig.~\ref{fig:Fig3}(a) until (excluded) the ${\rm LiH}_2$ Hamiltonian. Since the worst-case scaling of $r$ is exponential in the number of vertices \cite{moon1965cliques}, resorting to all maximal cliques is not an option when $n$ is large. We thus developed an algorithm that spans all vertices and, for each, finds a user-specified number of maximal cliques (see Sec.~\ref{sec:algorithm} and \cite{github} for more informations). When building each maximal clique, this algorithm favours vertices with more edges and higher contributions to the cost function $\costF$ in Eq.~\eqref{eq:gen_error_cov}. This creates a bias towards larger cliques, that in the context of measuring a quantum observable are highly desirable. In fact, we have tested {\algo} when using either all, or a subset of $r \lesssim n$ (large) maximal cliques. For the examples we considered, the errors yielded in these cases differed by $15\%$ at most.

The values reported in Fig.~\ref{fig:Fig3}(a-b) for the LDF method have been determined with an algorithm based on Ref.~\cite{jena2019pauli} and integrated with the framework developed in this work. We first find a cover of the graph such that each vertex is contained in exactly one clique. To find this cover, we also prioritise larger cliques and vertices with higher $\costF$ contributions, as we did for the maximal cliques' algorithm described in the previous paragraph. Once a cover is found, we use the BF subroutine to allocate all measurements, without ever updating the covariance matrix $\tilde{\mathcal{Q}}$, i.e., no adaptive features are used for the LDF method. We then perform the measurements, and resorting to the sample average we obtain $R=40$ [$R=100$] values $\tilde{O}_j$ that are used in Eq.~\eqref{eq:values_in_table} to determine the numbers reported in Fig.~\ref{fig:Fig3}(a) and \ref{fig:Fig3}(c) [Fig.~\ref{fig:Fig3}(b)].

The numerical results from the LDF method reported in Sec.~\ref{sec:results} are lower if compared to the ones reported in Refs.~\cite{jena2019pauli,verteletskyi2020measurement,hadfield2020measurements,Bujiao2021,Huang2021Efficient}. 
This is a consequence of two facts. First, for each graph cover, our measurement allocation is optimal (as can be demonstrated by following the procedure in App.~\ref{app:overlap}). 
Second, we allow for general commutation relations between PS, i.e., we group together two PS that commute, but do not necessarily bitwise commute. Indeed, when we restrict our LDF to bitwise commutation relations (as in App.~\ref{app:bitwise_comp}), it yields results that are similar to the ones reported for the LDF algorithm in Refs.~\cite{jena2019pauli,verteletskyi2020measurement,hadfield2020measurements,Bujiao2021,Huang2021Efficient}.

Our version of the LDF is representative of other measurement protocols (such as the ones in Refs.~\cite{jena2019pauli,hamamura2020efficient,Yen2019Measuring,Izmaylov2019,Gokhale2019}) that also allow for non-bitwise commuting relations between PS. In fact, prior to this work two criteria were commonly used to collect PS. First, the magnitude of their coefficients and second, the total number of resulting groups. For building the groups of PS to be measured together, our LDF protocol employs the expected contributions of the PS to the error, that is available via Eqs.~\eqref{eq:gen_error} and results in a strategy that is similar to the one used by {\algo} (described in Sec.~\ref{sec:algorithm} and \cite{github}). 
Since having less groups and collecting together PS with large coefficients are highly correlated with having lower estimation errors, we expect the algorithms in Refs.~\cite{jena2019pauli,hamamura2020efficient,Yen2019Measuring,Izmaylov2019,Gokhale2019} to achieve errors that are comparable to the numerical LDF results reported in Sec.~\ref{sec:results}.

In all panels of Fig.~\ref{fig:Fig4}, we report averaged values of $\costF$ with their statistical errors [in parentheses in (a) and as error bars in (b-e)] and $R=25$ [see above discussion about Fig.~\ref{fig:Fig3}(c)]. The same chemistry Hamiltonians previously utilized in Fig.~\ref{fig:Fig3} have been used in Fig.~\ref{fig:Fig4}(a), while we resorted to the family of lattice models introduced in the main text and App.~\ref{app:lattice_model} for Fig.~\ref{fig:Fig4}(b-d). In (a), (b) and (d), the settings chosen for estimating $\tilde{\mathcal{Q}}$ are the same ones used in Fig.~\ref{fig:Fig3}. For (c) and (e) the covariance matrix is initialized to the exact values, as explained at the bottom of Sec.~\ref{app:sec:our_algo_features}.

For all numerical values in Fig.~\ref{fig:Fig3} where the ML subroutine has been utilized, we employed a stochastic gradient descent optimization algorithm \cite{Kingma:2014aa} for the cost function $\costF$ minimization. This algorithm's performance depends on hyperparameters \cite{github,hoyer2019neural,Kingma:2014aa} -- the learning rate in particular -- that have not been optimized to keep low computational requirements. On the other hand, the ML subroutine in Fig.~\ref{fig:Fig4} employs the Limited-memory BFGS optimizer \cite{liu1989limited} that, belonging to the family of quasi-Newton methods \cite{gill1972quasi}, does not require choosing the learning rate. 


%
\section{Bitwise commutation comparison}\label{app:bitwise_comp}
\begin{figure}
    \centering
    \includegraphics[width=\columnwidth]{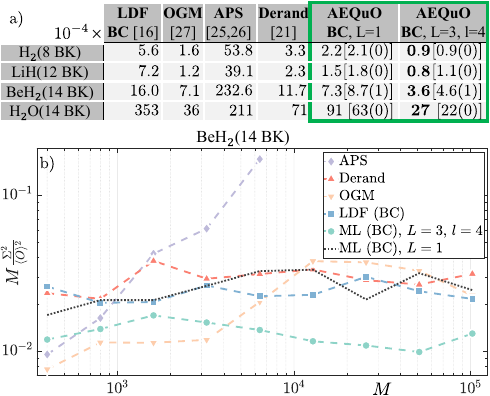}
    \caption{
    (a): Errors obtained with LDF \cite{jena2019pauli}, OGM \cite{Bujiao2021}, APS \cite{hadfield2020measurements,hadfield2021adaptive}, Derand \cite{Huang2021Efficient}, and {\algo} with the ML subroutine and $L=1$ or $L=3$ and $l=4$. We consider chemistry Hamiltonians in the BK encoding \cite{Cao2019Quantum}. Values (in bold the lowest) are variances $\Sigma^2$ with $\Sigma$ as in Eq.~\eqref{eq:values_in_table}. In square brackets, we report averages of $\costF$ 
    with their statistical error. The input is the ground state, $R=25$, $M=10^4$, and values are rescaled by $10^{-4}$. 
    (b): Relative errors $M \Sigma^2/\langle O \rangle^2$ 
    as a function of $M$ obtained considering the $\rm{BeH}_2$ Hamiltonian with BK encoding and $R=10^2$. In both panels, {\algo} and the LDF \cite{jena2019pauli} protocol are restricted to bitwise commutation (BC) relations.
    }
    \label{fig:FigS0}
\end{figure}

In this section, we compare {\algo} and our LDF protocols both restricted to bitwise commutation (BC) relations with the APS \cite{hadfield2020measurements,hadfield2021adaptive}, the Derand \cite{Huang2021Efficient} and the OGM \cite{Bujiao2021} methods. In fact, as explained in Sec.~\ref{sec:algorithm}, simultaneous measurements of non-bitwise commuting PS require entangling gates that, in the context of NISQ devices, generate errors that can jeopardize the result. As such, it is important to determine the advantage resulting from the adaptive nature of {\algo} alone. 

In Fig.~\ref{fig:FigS0} we present analogous numerical results as in Fig.~\ref{fig:Fig3}(a-b), with the additional constraint that all measurement schemes are restricted to bitwise commutation relations. From panel (a), it is possible to conclude that for the considered chemistry Hamiltonians the adaptive nature of {\algo} does provide a consistent advantage over all other schemes. Except for the water molecule, non-adaptive ($L=1$, black dotted line) {\algo}, the OGM, and the Derand protocols yield results that are within statistical fluctuations of one another (in agreement with Fig.~\ref{fig:FigS0}(b) and Figs.~\ref{fig:Fig3} and \ref{fig:Fig4}). Furthermore, while the LDF protocol in Fig.~\ref{fig:Fig3}(a-b) was yielding results that, in several instances, were better than the OGM and the Derand, here the LDF always performs comparably or worse. We thus conclude that its advantage came from non-bitwise commutation between PS that, in this section, is not allowed.

The case of $\rm{H}_2 \rm{O}$ in Fig.~\ref{fig:FigS0}(a) presents an interesting feature. The OGM, while performing worse than adaptive ($L=3$, $l=4$) {\algo}, outperforms all other approaches. After a careful analysis, we identified the reason of this advantage in the fact that the OGM protocol, for small measurement budgets, does not measure the input observable. Instead, it removes PS with small coefficients from the Hamiltonian \cite{Bujiao2021}. These terms, that are considered and measured by {\algo}, the LDF and the Derand protocols, introduce extra statistical fluctuations that increase the error. This is supported by Fig.~\ref{fig:FigS0}(b), where it is evident that the OGM (yellow downwards triangles) asymptotically reaches (within statistical fluctuations) the Derand (red upwards triangles) and non-adaptive {\algo} ($L=1$, black dotted line) for large values of $M$, when all PS are considered by OGM.

To confirm that the OGM's enhanced performance for small $M$ comes from neglecting PS that bare extra statistical errors, we investigated the performances of all protocols with different input states. For small values of $M$, we have found several instances in which the OGM yields results that are comparable to the Derand and non-adaptive ($L=1$) {\algo}.

\section{Lattice Hamiltonians}
\label{app:lattice_model}
\begin{figure}
    \centering
    \includegraphics[width=\columnwidth]{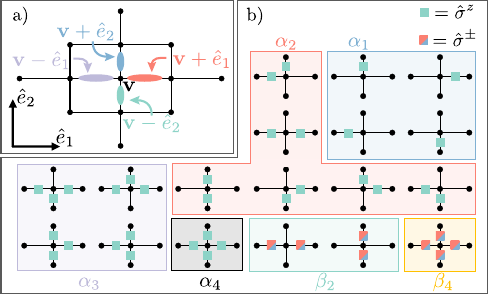}
    \caption{
    (a): Example of a lattice used in Fig.~\ref{fig:Fig4}(b-d). Here, we have $D=2$ dimensions and $d=16$. Vertices are indicated by their coordinates with vectors $\mathbf{v}$. Qubits lie on edges (coloured ellipses), and are identified by $\mathbf{v} \pm \hat{e}_i$ $(i=1,\dots,D)$ with $\hat{e}_i$ being unit versors defining the $D$-dimensional Cartesian plane. (b): All terms in the Hamiltonian describing the $D=2$ dimensions model with $d=4$ qubits. Green squares indicate $\hat{\sigma}^z$ interactions, while red/blue ones the flip-flop terms. We grouped together all interactions characterized by the same coefficients [$\alpha_k$ and $\beta_{2k}$ in Eq.~\eqref{eq:lattice_ham}].}
    \label{fig:FigS1}
\end{figure}
For deriving the data in Fig.~\ref{fig:Fig4}(b-d), we considered different $2$- and $3$D lattice models (see bottom of the figure), described by the Hamiltonians
%
\begin{equation}\label{eq:lattice_ham}
\begin{split}
    \mathcal{H} = & \sum_{\mathbf{v},i} \Bigg[ \sum_{k=1}^{2 D} \frac{\alpha_{k}}{2} \prod_{
    \substack{ S \subseteq \lbrace \hat{\mathbf{e}}, -\hat{\mathbf{e}} \rbrace
    \\ 
    \hat{i} \in S}
    }^{\lvert S \rvert = k} \hat{\sigma}_{\mathbf{v}+\hat{i}}^z +
     \\
    & \sum_{k=1}^{D} \beta_{2k} \prod_{
    \substack{ S \subseteq \lbrace \hat{\mathbf{e}} \rbrace
    \\ 
    \hat{i} \in S}
    }^{\lvert S \rvert = k} \left( \hat{\sigma}_{\mathbf{v}+\hat{i}}^+ \hat{\sigma}_{\mathbf{v}-\hat{i}}^- + \hat{\sigma}_{\mathbf{v}+\hat{i}}^- \hat{\sigma}_{\mathbf{v}-\hat{i}}^+ \right) \Bigg].
\end{split}
\end{equation}
To better explain this operator, we refer to Fig.~\ref{fig:FigS1}. In panel (a), we show an example of a lattice in $D=2$ dimensions with $d=16$ qubits lying on the edges. Vertices are indicated by vectors $\mathbf{v}$ and the Cartesian coordinate frame is characterized by versors $\lbrace \hat{\mathbf{e}} \rbrace = \lbrace \hat{e}_1,\dots,\hat{e}_{D} \rbrace$. It follows that each qubit can be identified by $\mathbf{v} \pm \hat{e}_{j}$ ($j=1,\dots,D$), where $\mathbf{v}$ is the vertex from which it can be reached by one of the versors (see Fig.~\ref{fig:Fig1}). In Eq.~\eqref{eq:lattice_ham}, $\alpha_k$ and $\beta_{2k}$ are real constants, and the products run over all possible combinations of versors ($S$ are subsets of $k$ oriented versors). The raising and lowering operators are indicated with $\hat{\sigma}^\pm = (\hat{\sigma}^x \pm i \hat{\sigma}^y)/2$. In Fig.~\ref{fig:FigS1}(b), all elements of the Hamiltonian in Eq.~\eqref{eq:lattice_ham} are represented for a $D=2$ dimensional lattice with $d=4$ qubits.

This class of physical models, describing many-body, next-neighbouring interacting spin-$1/2$ particles on a lattice is a generalization of the Hubbard model \cite{essler2005one} with hopping multi-particle terms and energy shifts depending on the particles' state. For determining the data points in Fig.~\ref{fig:Fig4}(b-e), we set $\alpha_k = 1/k$ and $\beta_{2k} = 1/(2k)$.

\printbibliography

\end{document}